\documentclass[english,nofootinbib,superscriptaddress,groupedaddress,aps,12pts]{revtex4}
\usepackage{color}
\usepackage{refstyle}
\usepackage{float}
\usepackage{amsthm}
\usepackage{amsmath}
\usepackage{amssymb}
\usepackage{graphicx}
\usepackage{esint}
\usepackage{xargs}[2008/03/08]
\usepackage[unicode=true,bookmarks=true,bookmarksnumbered=false,bookmarksopen=false, breaklinks=false,pdfborder={0 0 1},backref=false,colorlinks=true]{hyperref}
\hypersetup{pdftitle={Geometric Entropy}}
\usepackage{breakurl}
\usepackage{epstopdf}

\makeatletter

\AtBeginDocument{\providecommand\figref[1]{\ref{fig:#1}}}
\AtBeginDocument{\providecommand\eqref[1]{\ref{eq:#1}}}
\RS@ifundefined{subref}
  {\def\RSsubtxt{section~}\newref{sub}{name = \RSsubtxt}}
  {}
\RS@ifundefined{thmref}
  {\def\RSthmtxt{theorem~}\newref{thm}{name = \RSthmtxt}}
  {}
\RS@ifundefined{lemref}
  {\def\RSlemtxt{lemma~}\newref{lem}{name = \RSlemtxt}}
  {}

\@ifundefined{textcolor}{}
{%
 \definecolor{BLACK}{gray}{0}
 \definecolor{WHITE}{gray}{1}
 \definecolor{RED}{rgb}{1,0,0}
 \definecolor{GREEN}{rgb}{0,1,0}
 \definecolor{BLUE}{rgb}{0,0,1}
 \definecolor{CYAN}{cmyk}{1,0,0,0}
 \definecolor{MAGENTA}{cmyk}{0,1,0,0}
 \definecolor{YELLOW}{cmyk}{0,0,1,0}
}

\usepackage{amssymb}
\usepackage{amstext}
\usepackage{amscd}
\usepackage{amsthm}
\usepackage{amsopn}
\usepackage{amsmath}
\usepackage{amsbsy}
\usepackage{indentfirst}
\usepackage{mathrsfs}
\usepackage{color}
\usepackage{xcolor}
\usepackage{psfrag}
\usepackage{ulem}

\newcommand{\eveq}{\widehat{=}}
\newcommand{\vac}{\left|0\right>}
\newcommand{\Rvac}{\left\langle0\right|}
\newcommand{\RN}{{\mathcal{R}_N}}

\newcommand{\too}{\longrightarrow}
\newcommand{\Cal}[1]{{\cal #1}}
\newcommand{\be}{\begin{equation}}
\newcommand{\ee}{\end{equation}}
\newcommand{\bea}{\begin{eqnarray}}
\newcommand{\eea}{\end{eqnarray}}
\newcommand{\p}{\partial}

\newcommand{\ket}[1]{\left| #1 \right\rangle}
\newcommand{\bra}[1]{\left\langle #1 \right|}
\newcommand{\ind}[1]{\begin{footnotesize}\mbox{#1}\end{footnotesize}}
\newcommand{\vm}[1]{\left\langle #1 \right\rangle}
\newcommand{\VM}[1]{\big\langle #1 \big\rangle}

\newcommand{\f}{\textsc{f}}

\newcommand{\XXX}[1]{}

\renewcommand{\tilde}[1]{\widetilde{#1}}

\newcommand{\tr}{\mbox{Tr}}

\newcommand{\Hilb}{\mathscr{H}}

\newcommand{\hc}{\mbox{h.c.}}

\newcommand{\TB}{T_{\textsc{b}}}

\newcommand{\EB}[1]{\color{RED}#1\color{BLACK}}
\newcommand{\rmEB}[1]{\color{BLUE}\sout{#1}\color{BLACK}}

\renewcommand{\EB}[1]{#1}
\renewcommand{\rmEB}[1]{}

\newcommand{\Sentang}{\Cal S}
\newcommand{\HB}{H_\textsc{b}}
\newcommand{\SB}{S_\textsc{b}}
\newcommand{\tw}{\Phi}
\newcommand{\ttw}{\tilde\tw}
\newcommand{\iorb}{\!\! \mbox{\begin{scriptsize} orb\end{scriptsize}}}
\newcommand{\res}[2]{\mbox{Res}_{#1}\big[#2\big]}
\newcommand{\frn}{f_{N}^{\hphantom{\int}}}
\newcommand{\fS}{f_{\Sentang}^{\hphantom{\int}}}
\newcommand{\eemph}[1]{{\it #1}}
\newcommand{\RNIR}{R_N^{\textsc{ir}}}
\newcommand{\SIR}{\Cal S^{\textsc{ir}}}
\newcommand{\eimp}{^{\mbox{\begin{scriptsize}{imp}\end{scriptsize}}}}
\newcommand{\Zimp}{\mathcal{Z}\eimp}

\@ifundefined{showcaptionsetup}{}{%
 \PassOptionsToPackage{caption=false}{subfig}}
\usepackage{subfig}
\makeatother

\begin{document}

\global\long\def\lorbivac{\left|\mbox{orb}\right\rangle }
\global\long\def\rorbivac{\left\langle \mbox{orb}\right|}

\global\long\def\twistedvac{\left|\sigma\right\rangle }
\global\long\def\ktwistedvac{\left|\sigma^{(k)}\right\rangle }

\global\long\def\CN{{{\cal C}_{N}}}

\newcommandx\VMRN[1][usedefault, addprefix=\global, 1=]{\VM{#1}_{\RN}}

\newcommandx\VMORB[1][usedefault, addprefix=\global, 1=]{\frac{\VM{#1\Phi(u)\tilde{\Phi}(v)}}{\VM{\Phi(u)\tilde{\Phi}(v)}}}

\newcommandx\VMsurface[2][usedefault, addprefix=\global, 1=]{\VM{#2}_{#1}}

\title{Infra-red expansion of entanglement entropy in the Interacting
Resonant Level Model}
\author{L. Freton}
\affiliation{Laboratoire Mat\'eriaux et Ph\'enom\`enes Quantiques,
Universit\'e Paris  Diderot, CNRS UMR 7162, 75013 Paris, France}
\author{E. Boulat}
\affiliation{Laboratoire Mat\'eriaux et Ph\'enom\`enes Quantiques,
Universit\'e Paris  Diderot, CNRS UMR 7162, 75013 Paris, France}
\author{H. Saleur}
\affiliation{Institut de Physique Th\'eorique, CEA, IPhT and CNRS, URA2306, Gif Sur Yvette, F-91191}
\affiliation{Department of Physics, University of Southern California, Los Angeles, CA 90089-0484}

\begin{abstract}
In this paper we develop a method to describe perturbatively the entanglement
entropy in a simple impurity model, the interacting resonant level
model (IRLM), at low energy (i.e. in the strong coupling regime). We use
integrability results for the Kondo model to describe the infra-red
fixed point, conformal field theory techniques initially developped
by Cardy and Calabrese and a quantization scheme that allows one to
compute exactly Renyi entropies  at arbitrary order in $1/\TB$ in principle, even
when the system size or the temperature is finite. We show that those universal quantities at arbitrary interaction parameter in the strong coupling regime are very well approximated by the same quantities 
in the free fermion system in the case of attractive Coulomb interaction, whereas a strong dependence on the interaction appears in the case of repulsive interaction.
\end{abstract}
\maketitle

\section{Introduction}

Entanglement is a property allowed by quantum mechanics that describes the fact that generically, a quantum state of a system consisting of several subparts cannot be written as a product of states of the subparts. One way to characterize, at zero temperature, this fascinating property of quantum systems, is to introduce the so-called entanglement entropy. Since it was first considered, this quantity has found many applications, ranging from the study of blakholes \cite{blackholes,holzhey94}, to modern methods of simulation of quantum systems \cite{dmrg}.

The study of quantum impurity problems from the point of view  of entanglement entropy has led to many new insights and puzzles, both in \cite{ALS09} and out of equilibrium situations \cite{KL09,SFRKH10}. In these problems, one typically considers a 1D gapless bath coupled to a localized degree of freedom, and the questions of interest concern the entanglement of the two halves separated by the impurity \cite{L04,P05,ZPW06,goldstein11} , the entanglement of the impurity with the bath \cite{CK03,SCLA07}, or the entanglement of a region containing the impurity with the rest of the system \cite{FFN07}. In all those situations, one is dealing with a bipartite entanglement entropy (associated to a partition of the system in two pieces A and B)  that characterizes to what extent the groundstate of the full system can be factorized into states for the subsystems A and B. Usually, the coupling of the impurity to the bath leads to a renormalization group flow, so the entanglement entropy exhibits different crossovers, and cannot be studied by the powerful methods of conformal field theory \cite{CC09}.

While numerical results are available for several examples of the foregoing problems, analytical results are very few \footnote{A remarkable exception is the calculation in \cite{KL09} of the {\sl rate} of entanglement entropy in the steady, out of equilibrium state of two wires connected by a resonant level, i.e. without interactions.}. Several difficulties are compounded here. One is the complexity of the entanglement entropy, which usually does not behave simply under the various Bethe ansatz and integrable quantum field theory  tricks. Another, maybe less immediately obvious, comes from the fact the entanglement entropy, being a $T=0$ quantity, is very hard to access perturbatively because of Infra Red (IR) divergencies - its behavior in perturbation theory is in fact reminiscent of the screening cloud in the Kondo problem \cite{BA96,A09}. Like for the Kondo model,  the models we are interested in also  involve flow to strong coupling at low energy: no matter how small the bare impurity/bath coupling $\gamma$ is, the asymptotically low energy states (below some scale $\TB(\gamma)$ that goes to zero when $\gamma\to 0$) of the theory are qualitatively and deeply affected by the coupling. In more technical, RG terms, the interaction between the impurity and the bath is a (maybe marginally) {\sl relevant operator}, of scaling dimension $D\in [0,1]$. 
The RG flow is  characterized by an  energy scale $\TB$ (that coincides with the Kondo  temperature in the case of the Kondo model), which  marks the crossover between the weakly interacting regime, or ultraviolet (UV) regime, and the strong coupling regime, or infrared (IR) regime. 
By analogy with the usual thermodynamics quantities - such as the impurity entropy \cite{AL91} - one expects the entanglement and Renyi entropies of the impurity, to bear a universal form and depend only on dimensionless parameters in the low energy (with respect to the cutoff - typically the bandwidth) limit :
\begin{equation}
R_N\eimp = \frac{R_N}{ R_N^{\ind{bulk}}} = \frn \left(\frac{v_\f}{L\TB},\frac{v_\f}{L_0\TB},\frac{1}{\beta\TB},...\right)\qquad \Sentang\eimp = \Sentang-\Sentang^{\ind{bulk}} = \fS \left(\frac{v_\f}{L\TB},\frac{v_\f}{L_0\TB},\frac{1}{\beta\TB},...\right)
\label{scalfunctions}
\end{equation}
where $L$ the length of subsystem $A$, $L_0$ the size of the system,  $\beta^{-1}$ is the temperature, ..., and where by "bulk" quantities we mean  those of the system with no impurity\footnote{Depending on the geometry, this system with no impurity may have a boundary.}. Scaling analysis meanwhile shows that $\TB\propto \gamma^{1/1-D}$. While the validity of the form (\ref{scalfunctions}) seems well established numerically \cite{SCLA07, ALS09} at least in some cases, the natural attempt to calculate the universal functions perturbatively in $\gamma$  fails: indeed, while it is easy to set up the calculation and bring it down to the evaluation of explicit integrals, all orders turn out to be infinite \cite{BFSVP}, despite the presence of a natural IR cutoff - the length $L$ of the subsystem.  While these divergences  simply mean that the scaling functions are not analytic in $\gamma$, they leave us  facing  the practical problem of evaluating a non perturbative quantity in a particularly difficult setup involving replicas etc. The present  paper and its sequels are devoted to making progress on this important question, with the ultimate goal of computing entanglement rates in out of equilibrium, interacting situations.

The impurity models we are going to consider have in common that 
the impurity degree of freedom is a two-state system (occupied or empty dot in the case of the IRLM, spin $\frac{1}{2}$ for the Kondo model). At the IR fixed point, the impurity will end up being "absorbed by the bath", the physical picture in a lattice regularization of the model being that it forms a singlet with the bath site neighboring the impurity \cite{affleckreview}. In a continuum model, the hybridization of the impurity
with the bath should be rather thought of as occurring in a region of size $\lambda_{\textsc{b}}=v_\f/\TB$, the so-called "Kondo cloud" in  the Kondo model. At zero temperature, one thus expects that the impurity has \eemph{no} effect on the entropies  $R_N,\Sentang$ when the partitioning of the system is done well beyond the impurity cloud (i.e. when the cloud is well inside region A, $L\gg \lambda_B$), resulting in the simple limits $\frn(0,0,0)=1$ and $\fS(0,0,0)=0$. Another simple limit at zero temperature is that of a disconnected impurity $\gamma\to 0^+$ or $\TB\to 0$, where one has 
\footnote{Note that the point $\gamma=0$ is singular since then the impurity density matrix is totally arbitrary, not being specified by the contact with the bath.}
$\frn(\infty,x,0)=2^{1-N}$ and $\fS(\infty,x,0)=\ln  2$. The entanglement entropy thus has the \emph{same} limits in the UV and IR regime as does the boundary entropy $s=\ln (g)$ (in this general formula $g$ is the "universal non-integer groundstate degeneracy") introduced by Affleck and Ludwig \cite{affleckludwig1,affleckludwig2} when varying the temperature between $\beta^{-1}=\infty$ and $\beta^{-1}=0$.
In between those two simple limits lies a crossover regime where the interactions play at full and shape the crossover functions. It is probably tempting to speculate that, since  they interpolate between the same values, the entanglement and boundary entropies could be the same functions, after  formally replacing the length of region A by the inverse temperature. We shall see below that this is definitely not the case, and that entanglement and boundary entropies behave very differently, as a function of the interactions in particular. 

\begin{figure}[ht]
\begin{centering}
\includegraphics[height=4.5cm]{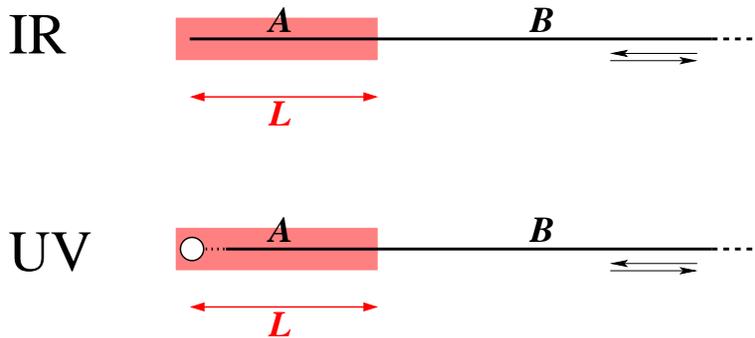}
\par\end{centering}
\caption{Illustration of the low energy or long distance (IR) and high energy or short distance (UV) limits of the Kondo model, after reduction to the s-wave channel. In the low energy limit, the impurity spin is swallowed by the bath, while it is decoupled at  high energy (denoted by a white circle). The black arrows denote right- and left-moving fields. $A$ and $B$ denote the partition of the system that is used to define the bipartite entanglement.}
\label{fig1}
\end{figure}

The Kondo situation is illustrated on figure \ref{fig1} where we have represented the  spin   absorbed by the bath  in the low energy (long distance) ("infrared" or IR) limit, and the spin decoupled from the bath  at high energy (short distance) ("ultraviolet" or UV). It is well known in these two, conformal  limits, that the entanglement entropy of the region of length $L$ with the rest of the system behaves as\footnote{Recall also that the entanglement entropy of a region of length $L$ sitting not at the edge but in the bulk of an infinite system takes the related form ${\cal S}={c\over 3}\ln{L\over a}+s_1$. }

\begin{equation}
{\cal S}={c\over 6}\ln{2L\over a}+{s_1\over 2}+\ln g
\end{equation}
(where $a$ is a cutoff length scale,  and $s_1$ a non universal constant) so ${\cal S}\eimp$ interpolates between $\ln 2$ and $0$. 

The  Kondo model (or the IRLM also considered below) being integrable, it is natural to wonder about determining the entanglement exactly using Bethe ansatz type techniques. This however remains a seemingly unreachable goal, despite  progress such as in \cite{FIKT10}. The next best option is  to obtain what are usually extraordinarily good approximations using the form factors technique, directly in the continuum, scaling theory, after taking the appropriate ``massless limit'' of the usual massive construction \cite{CCD07}. This will in fact be discussed in our next paper. Meanwhile, there is one last tool we can try to put to use, following the calculation to lowest order performed in \cite{SCLA07}: the expansion around the strong coupling fixed point. Such an expansion is bound to be better behaved as far as IR divergences are concerned, since one perturbs by irrelevant operators, whose correlation functions decay sufficiently fast at infinity to render all integrals convergent at large distance. Indeed, the lowest order calculation proposed in  \cite{SCLA07} relies on the effective Fermi Liquid theory for the vicinity of the Kondo strong coupling fixed point, and boils down, technically, to the calculation of a single integral of the one point function of the stress energy tensor (see below), a perfectly well defined procedure, leading to a result that depends only on $\TB,L$ as required. This procedure is however bound to fail in general, beyond the first order, since there are now strong UV divergences, and perturbation by an irrelevant operator does not determine a renormalizable field theory.  However, in the integrable case, the existence of an infinity of conserved quantities does, in fact, provide a full control of the low energy hamiltonian: the necessarily infinite number of counter terms are all explicitly known \cite{LS99}, together with a well defined (analytical) regularization procedure.  This means that, while we do not know how to calculate the scaling functions  at small coupling (since they are not perturbative), we have, in principle, a tool to determine them perturbatively at large coupling. Of course, only a few orders are technically manageable, but this is enough to gain an understanding of the scaling functions, and answer important qualitative questions.

The paper is organized as follows:
In section \ref{model}, we introduce the impurity model and its representation in the ultraviolet and infrared limits.
In section \ref{method}, we present the method of infrared perturbation theory in the replicated theory needed to compute the entropies. We then turn to the actual calculation of the entropies in an infinite size system
in section \ref{infinitesize}: our result for the entanglement entropy is summarized in Eqs. (\ref{result:infinite},\ref{result:infinite:imp}).  Section \ref{finitesize} treats the case of a  system of finite  size $2L_0$, our results being summarized in Eqs. (\ref{result:finite},\ref{result:finite:imp}).

\section{Model}
\label{model}

\subsection{Interacting Resonant Level Model}

We start in fact not with the Kondo hamiltonian, but with the interacting resonant level model  (IRLM). This  is a simple impurity model that describes a tunnel junction
between a localized resonant level at $x=0$ and two baths of free,
spinless electrons. We note here that  the model and all the manipulations below can be generalized to an arbitrary  number of baths, 
but this will not change qualitatively the results very much. Later, we will discuss briefly  the one-wire case. 

The IRLM also includes  a Coulomb interaction (that is also called "excitonic" interaction)
between the level and the baths. After a mode expansion in the baths
and a linearization around the Fermi points, we end up with an hamiltonian
$H=H_{0}+\HB$, where $H_{0}=\sum_{a=1,2}-i\int_{-\infty}^{\infty}dx\psi_{a}^{\dagger}\p_{x}\psi_{a}$
(we have unfolded the fermionic fields living on the half-infinite
lines into right-moving fields on the complete line). The boundary
interaction is given by
\begin{equation}
\begin{aligned}\HB= & \sum_{a=1,2}\gamma_{a}\psi_{a}(0)d^{\dagger}+\hc\\
 & +U\left(\psi_{1}^{\dagger}\psi_{1}+\psi_{2}^{\dagger}\psi_{2}\right)(0)(d^{\dagger}d-1/2).
\end{aligned}
\end{equation}
It is convenient to use spin-1/2 operators $d^{\dagger}=\eta_{d}S^{+}$
and $d^{\dagger}d=S_{z}+1/2$ to represent the impurity ($\eta_{d}$
is a Majorana fermion). The tunneling anisotropy is parametrized with
$\gamma\sqrt{2}e^{i\theta/2}=\gamma_{1}+i\gamma_{2}$ and we consider
even and odd fermions defined as a rotation of the original fermions
\begin{equation}
\begin{pmatrix}\psi_{+}\\
\psi_{-}
\end{pmatrix}=\begin{pmatrix}\cos\frac{\theta}{2} & \sin\frac{\theta}{2}\\
-\sin\frac{\theta}{2} & \cos\frac{\theta}{2}
\end{pmatrix}\begin{pmatrix}\psi_{1}\\
\psi_{2}
\end{pmatrix}
\end{equation}
We take the bosonized form of this model with $\psi_{a}(x)=\eta_{a}:e^{i\sqrt{4\pi}\varphi_{a}(x)}:/\sqrt{2\pi}$,
with $a=1,2,\pm$. Then $(\kappa_{a}=\eta_{a}\eta_{d})$ 
\begin{align}
H_{0}\left[\varphi_{a}\right] & =\sum_{a=1,2}\frac{1}{2}\int_{-\infty}^{\infty}dx\left(\p_{x}\varphi_{a}\right)^{2}=H_{0}\left[\varphi_{\pm}\right]
\end{align}
\begin{equation}
\begin{aligned}\HB & =\frac{\gamma}{\sqrt{\pi}}\left[\kappa_{+}:e^{i\sqrt{4\pi}\varphi_{+}(0)}:S^{+}+\hc\right]\\
 & +\frac{U_{b}}{\sqrt{\pi}}\left(\p_{x}\varphi_{+}+\p_{x}\varphi_{-}\right)(0)S_{z}
\end{aligned}
\end{equation}
where $U_{b}=g(U)$ is a function of the capacitive term for fermions.
This function depends on the regularizing scheme chosen to bosonize
the initial model \cite{schillerandrei}.

\subsection{UV hamiltonian : anistropic Kondo model}

As in \cite{boulat_exact_2008} one can absorb the interaction
via a unitary transformation ${\cal U}=:e^{i\alpha\left(\varphi_{+}+\varphi_{-}\right)(0)S_{z}}:$
with $\alpha=U_{b}/\sqrt{\pi}$ that relates the model to the anisotropic
Kondo hamiltonian where the boson $\phi_{-}$ has decoupled
\begin{equation}
H=H_{0}\left[\phi_{\pm}\right]+\frac{\gamma}{\sqrt{\pi}}\kappa_{+}:e^{i\lambda\phi_{+}(0)}:S^{+}+\hc
\end{equation}
where $\lambda^{2}=\frac{2}{\pi}\left(U_{b}-\pi\right)^{2}+2\pi$ and
the bosons $\phi_{\pm}$ are again a rotation of the bosons $\varphi_{\pm}$
\begin{equation}
\begin{pmatrix}\phi_{+}\\
\phi_{-}
\end{pmatrix}=\frac{1}{\lambda}\begin{pmatrix}\frac{2\pi-U_{b}}{\sqrt{\pi}} & \frac{-U_{b}}{\sqrt{\pi}}\\
\frac{U_{b}}{\sqrt{\pi}} & \frac{2\pi-U_{b}}{\sqrt{\pi}}
\end{pmatrix}\begin{pmatrix}\varphi_{+}\\
\varphi_{-}
\end{pmatrix}
\end{equation}
the scaling dimension of the tunneling operator is $D=\lambda^{2}/8\pi$,
thus for $\lambda^{2}>8\pi$ the tunneling is irrelevant and the model
flows to its UV fixed point which consist of free right-moving bosons
living on the complete line.

In the following, we will be interested only in the interesting situation where the boundary term is relevant ($D<1$), and deeply influences the low-energy physics of the system by driving it to a strong coupling fixed point:  the model flows at low energy to the Kondo fixed point where the impurity
hybridizes with the wires and we get again free bosons with different
boundary conditions for $\phi_{+}$, namely $\phi_{+}(0^{+})=\phi_{+}(0^{-})+\lambda/4$,
whereas we have trivial boundary conditions for $\phi_{-}$ , that is
decoupled from the boundary. 
The renormalization group flow to strong coupling generates dynamically a new energy scale $\TB$, that marks the crossover between the UV fixed point ($\gamma=0$) and the IR fixed point.
All physical quantities are universal functions of $1/\beta\TB,1/L\TB$, etc.
This scale $\TB$ itself is of course non-universal, and it will depend
on $\gamma$ and $D$. A simple scaling argument leads to the power-law dependence $\TB \sim \gamma^{\frac{1}{1-D}}$.   

\subsection{The entanglement entropy}

\begin{figure}[ht]
\begin{centering}
\includegraphics[width=0.6\linewidth]{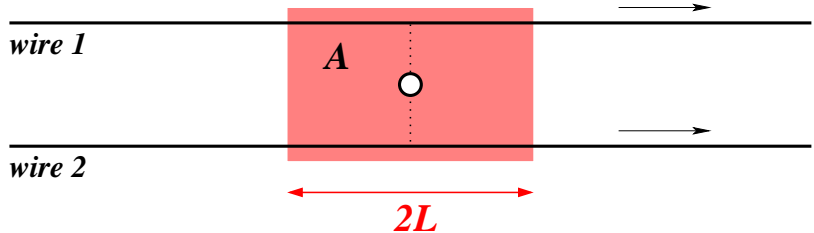}
\par\end{centering}
\caption{The IRLM after unfolding the wires to get only right movers.}
\label{fig2}
\end{figure}

\begin{figure}[ht]
\begin{centering}
\includegraphics[width=0.6\linewidth]{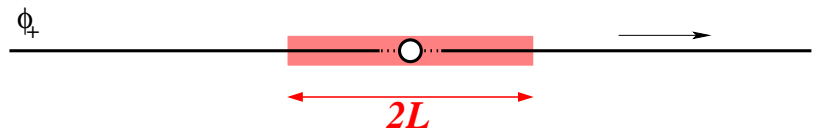}
\par\end{centering}
\caption{The $\phi_+$ degrees of freedom correspond to a single chiral wire coupled to the impurity.}
\label{fig3}
\end{figure}

The quantity we are initially interested in is the entanglement between a region (region A)  of size 
\footnote{The unfolding procedure doubles the spatial size of region A.}
 $2L$ centered around the resonant level and the rest of the system (region B). Unfolding gives us two chiral wires connected by the resonant level as illustrated on figure \ref{fig2}. After the unitary transformation, the entanglement can be decomposed into a sum of contributions from the $\phi_+$ and $\phi_-$ modes. Only the $\phi_+$ modes interact with the impurity as illustrated on figure \ref{fig3}. Finally, one may ``refold'' back the result to obtain exactly the Kondo geometry from figure \ref{fig1}. 

Note that the geometry we consider does not correspond to studying the entanglement between the two wires connected by the resonant level \cite{L04}. This situation is  somewhat more complicated, in particular because the problem then cannot be turned into a purely chiral one - we will discuss it elsewbere \cite{BFSVP}.

\subsection{Dual hamiltonian : approach of IR fixed point}

Since the boson $\phi_-$ decouples from the impurity, we are left with a single boson and  in the following we simplify notations and set $\phi_{+}\equiv\phi$.
Remarkably, as was shown by \cite{lesage_perturbation_1999}, due
to the integrability of the anisotropic Kondo hamiltonian, one has
a dual expression of the hamiltonian that is an expansion around
the IR hamiltonian $H_{0}^{IR}=\frac{1}{2}\int_{-\infty}^{\infty}dx\left(\p_{x}\phi\right)^{2}$
(altogether with its boundary conditions). The expansion takes the form of a series
of irrelevant boundary operators 
\begin{equation}
H=H_{0}^{IR}+\sum_{n\geq0}\frac{g_{2n+2}}{\TB^{2n+1}}{\cal O}_{2n+2}(0).
\label{eq:hamIR}
\end{equation}
The operators ${\cal O}_{2n+2}$ are expressed only in term of a (holomorphic)
deformed energy momentum tensor $\tilde{T}=T-i\sqrt{2\pi}\alpha\p^{2}\phi$,
with $T(z)=-2\pi:\p\phi^{2}:(z)$ and the coefficient $\alpha$ is related to the scaling dimension of the boundary perturbation via $\alpha=\frac{1-D}{\sqrt{D}}$. The deformed theory with energy momentum
tensor $\tilde{T}(z)$ has a modified central charge $\tilde{c}=1-6\alpha^2=1-6\frac{(1-D)^{2}}{D}$.
The operators ${\cal O}_{2n+2}$ are conserved quantities of the UV hamiltonian, they commute
with $H$ and between one another. The first terms read:
\begin{align}
{\cal O}_{2} & =\tilde{T}\\
{\cal O}_{4} & =(\tilde{T}^{2})\\
{\cal O}_{6} & =({\cal O}_{2}{\cal O}_{4})+\frac{\tilde{c}+2}{12}({\cal O}_{2}\EB{\p^2}{\cal O}_{2})
\end{align}
We also know the value of the coefficients $g_{2n+2}$, they read
\cite{lesage_perturbation_1999}
\begin{widetext}
\begin{align}
g_{2n+2}  = \frac{1}{2\pi}\; \EB{\frac{(-1)^{n+1}\left(D/\pi\right)^{n}}{(n+1/2)(n+1)!}}
  \times\left[\frac{\Gamma\left(\frac{D}{2(1-D)}\right)}{\Gamma\left(\frac{1}{2(1-D)}\right)}\right]^{2n+1}\frac{\Gamma\left(\frac{2n+1}{2(1-D)}\right)}{\Gamma\left(\frac{(2n+1)D}{2(1-D)}\right)}
\end{align}
\end{widetext}

Note that considering the IR fixed point amounts to taking the limit
$\TB\to\infty$, or equivalently to set all physical energy scales
(temperature, inverse size of the box $v_\f/L$,...) to $0$.

We are now able to develop a theory of perturbations in $1/\TB$ around
the IR fixed point for the Renyi entropies.


\section{Method}
\label{method}

\subsection{Renyi and entanglement entropies}
When a system is divided into two sub-systems $A$ and $B$ (more precisely when the Hilbert space
can be written as $\Hilb_A\otimes\Hilb_B$), one can define
 the Renyi entropy of order $N$ associated to the partition as:
 \begin{equation}
R_N\equiv R^{AB}_{N}=\tr_{A}\left(\rho_{A}\right)^{N}
\label{defRenyi}
\end{equation}
where $\rho_A$ is the reduced density matrix of subsystem $A$ obtained by tracing out $B$ degrees of freedom, $\rho_{A}=\tr_{B}\rho$ and the
partial traces $\tr_{X}(\cdot)=\sum_{\ket{\xi}\in\Hilb_X}\bra{\xi}\cdot\ket{\xi}$ take the trace on only a subpart $X$ of
the degrees of freedom. Note that Eq.(\ref{defRenyi}) makes sense (i.e. the sum converges absolutely)  for complex $N$ with $\mbox{Re}(N)>1$
The Renyi entropy has the property $R^{AB}_N=R^{BA}_N$ \cite{RA_B=R_BA}, hence the abreviation $R_N$ used in (\ref{defRenyi}).
The bipartite entanglement entropy between $A$ and $B$ is defined as 
\begin{equation}
\Sentang=\Sentang_{AB}=\tr_{A}\rho_{A}\ln\rho_{A}=\Sentang_{BA}
\label{defEntang}
\end{equation}
One can deduce the entanglement entropy from the knowledge of the Renyi entropies by using 
$\Sentang=-\lim_{N\to 1}\p_N R_N$. The replica trick is a way to compute the logarithm : if we take integer
powers of the reduced density matrix, the Renyi entropy can be interpreted (up to a factor) as the partition function of a big system consisting of $N$ (coupled) copies of the original theory. Then, one analytically continues the result to arbitrary $N\in\mathbb{C}$, $\mbox{Re}(N)>1$ to obtain $\Sentang$ (see Ref. \onlinecite{calabresecardy09} for considerations about the non-existence of ambiguities in the continuation -- essentially because $N=\infty$ is an accumulation point). Note that the replica trick


\subsection{The replica trick}

We now specify the partition $A/B$, and choose a  spatial division
of the system into a single finite interval and its complement, i.e.  $x\in A$ if $-L<x<L$ and $x\in B$ otherwise.
In the case where the initial full system has conformal symmetry, it was shown in \cite{calabresecardy04,calabresecardy09} that  the replicated theory bears a very elegant structure and can be described by a $\mathbb{Z}_N$ orbifold theory \cite{dixon87,cyclicorbifold97}. 

\begin{figure}[ht]
\begin{centering}

\subfloat[A matrix element of the full density matrix $\rho$]{
\begin{centering}
\includegraphics[width=5cm]{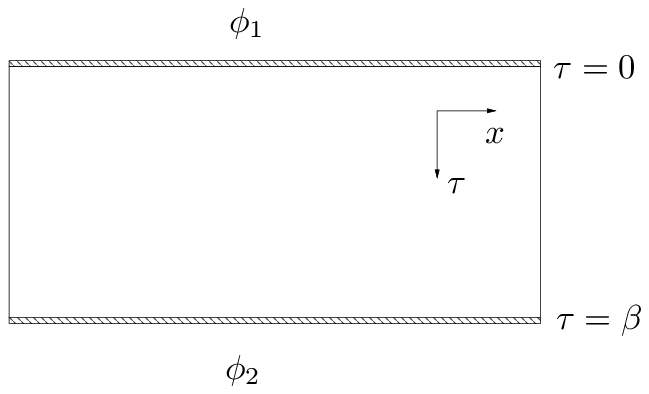}
\par\end{centering}
\centering{}
\label{fig:rho}
}\textcompwordmark{}

\subfloat[Reduced density matrix at finite temperature $\beta^{-1}$.]{
\begin{centering}
\includegraphics[width=7cm]{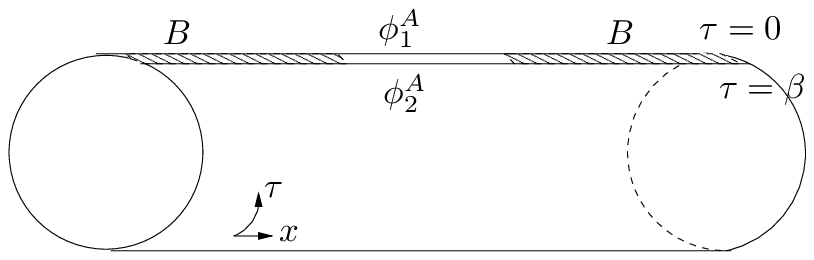}
\par\end{centering}
\centering{}
\label{fig:partialTraceFinite}
}\textcompwordmark{}
\subfloat[At zero temperature $\beta=\infty$.]{\centering{}
\includegraphics[width=7cm]{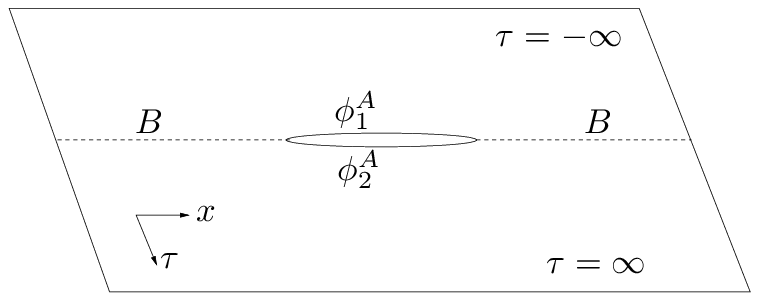}
\label{fig:partialtraceInfinite}
}
\par\end{centering}
\caption{
\small
Graphical representation of full (\figref{rho}) and reduced (\figref{partialTraceFinite,partialtraceInfinite})
 density matrix elements $\bra{\phi_1(x)}\rho_{(A)} \ket{\phi_2(x)}$. 
The fields $\phi_{1},\phi_{2}$ that label
the matrix element are the boundaries in the path integral formalism. 
In this formulation, the partial trace over $B$ amounts to gluing together
the boundaries in  region $x\in B$.
The fields $\phi$ propagate with the complete action $S$.
}
\end{figure}

To arrive at this, Calabrese and Cardy first  work with an Euclidian path integral representation of the full density matrix $\rho$ at temperature $\beta^{-1}$. If  the action depends on a collection
of fields : $S\left[\left\{ \phi(x,\tau)\right\} \right]$, a matrix
element of the density matrix $\rho$ is written:
\begin{equation}
\left\langle \left\{ \phi_{2}\right\} |\rho|\left\{ \phi_{1}\right\} \right\rangle =\frac{\int_{\phi(x,0)=\phi_{1}(x)}^{\phi(x,\beta)=\phi_{2}(x)}{\cal D}\phi\: e^{-S[\left\{ \phi\right\} ]}}{\int_{\phi(x,0)=\phi(x,\beta)}{\cal D}\phi\: e^{-S[\left\{ \phi\right\} ]}}\label{eq:rhopathintegral}
\end{equation}
which is represented graphically by \figref{rho}. Note that in our case, contrarily to \cite{calabresecardy04,calabresecardy09}, the action $S=S_0+S_{\textsc{b}}$ is not the free CFT action $S_0$.  

A matrix element of the reduced density matrix $\rho_{A}$ is labelled
by two spatial configurations of the field  $\phi^{A}_{1(2)}(x\in A)$ which live only in
the region $A$ : $\left\langle \phi_{2}^{A}|\rho_{A}|\phi_{1}^{A}\right\rangle $.
Since the partial trace amounts to take $\phi(x\in B,0)=\phi(x\in B,\beta)$
and to sum over the fields $\phi(x\in B)$, graphically this is represented
by a gluing of the boundaries at $\tau=0$ and $\tau=\beta$ in region $B$, leaving
a surface with a cut along $x\in[-L,L]$, over the lips of which the fields $\phi^A_{1(2)}(x)$ live.
This is shown at finite and zero temperature in \figref{partialTraceFinite}
and \figref{partialtraceInfinite}.

Performing the last trace over $A$ yields  a compact expression for the Renyi entropies $R_{N}$ that can be expressed as a ratio of partition functions:
\begin{eqnarray}
R_{N} & = & \frac{{\cal Z}_{N}}{({\cal Z}_{1})^{N}}\\
{\cal Z}_{N} & = & \int_{\RN(\beta)}{\cal D}\phi\; e^{-S[\phi]}\label{eq:renyi}.
\end{eqnarray}
Here $\Cal Z_N$ is the partition function of the theory living on the surface $\RN(\beta)$ that is obtained by gluing cyclically the $N$ open cuts of \figref{partialTraceFinite,partialtraceInfinite}, so that $\RN(\beta)$ is 
the $N$-sheeted Riemann surface $\RN$ when $\beta=\infty$, and a more complicated surface of higher genius at finite temperature.\\

\begin{figure}
\begin{centering}
\includegraphics[width=8cm]{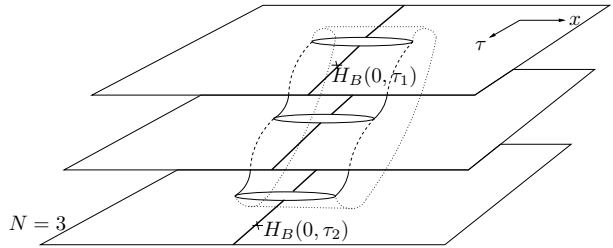}
\par\end{centering}
\caption{\small The $\RN$ geometry. The thick line at $x=0$ is the integration contours
of the boundary operators we insert. In this example we have inserted
two operators, coming from two different copies of the system.}

\label{fig:RNgeom}
\end{figure}

\paragraph{The $\RN$ geometry}
We now see the geometry in which we will have to compute correlators
of boundary operators in order to get the Renyi entropies. The matrix
$\rho_{A}^{N}$ is a fusion of the cylinders/planes via the "lips"
of the cuts, and taking the trace over region $B$ amounts to fuse
the last one with the first one. This gives the $\RN$ geometry, shown
in \figref{RNgeom}. In the path integral formalism this means that
we take $N$ copies of the system, we have a collection of fields
$\phi_{(j)}$ , $j=1\dots N$, and two inverted branching points,
one at $z=-iL$ and one at $z=iL$, so that $\phi_{(j)}(0^{+}-ix)=\phi_{(j-1)}(0^{-}-ix)$
for $-L<x<L$.

\subsection{Infra-red perturbation}

\subsubsection{Inserting operators on $\RN$}

\begin{figure}[h]
 \begin{centering}
\includegraphics[height=3.5cm]{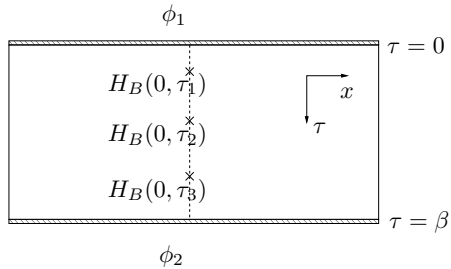}
\par\end{centering}

\caption{\small A matrix element of the full density matrix $\rho$, here at order $n=3$. The fields $\phi_{1},\phi_{2}$ that label
the matrix element are the boundaries in the path integral formalism. Propagation of the fields $\phi$ with the free action $S_{0}$ and insertion of boundary operators.}

\label{fig:rhoinsere}
\end{figure}

We want to make a perturbative expansion around the free action
$S_{0}$ : $S=S_{0}+\SB$ with $\SB=\int_{0}^{\beta}d\tau \HB(0,\tau) $
and we develop in the path integral (\ref{eq:rhopathintegral}) the
exponential 
\begin{equation}
\exp(-\SB)=\sum_{n=0}^{\infty}\int d\tau_{1}\dots \int d\tau_{n}\frac{(-1)^{n}}{n!}\prod_{i=1}^{n}\HB(0,\tau_{i}).
\end{equation}
Graphically a matrix element of $\rho$ is represented within this
expansion in \figref{rhoinsere}. 

 Expanding these path integrals around $H_{0}$,
one gets an expansion in powers of $1/\TB$ of the Renyi entropy
and thus of the entanglement entropy. The expansion is of the form
\begin{align}
R_{N} & =\int_{\RN}\mathcal{D}\varphi_{+}\mathcal{D}\varphi_{-}e^{-S}/\left(\int_{{\cal R}_{1}}\mathcal{D}\varphi_{+}\mathcal{D}\varphi_{-}e^{-S}\right)^{N}\label{eq:renyientropy}\\
 & =\RNIR\;
\frac{\VMRN[e^{-S_B}]}{\VM{e^{-S_B}}_{\mathcal{R}_1}}=\RNIR \frac{\Cal Z\eimp_N}{(\Cal Z\eimp)^N}
\label{eq:perturbDvt}
\end{align}
where $\RNIR$ is the unperturbed Renyi entropy, and is obtained from Eq. (\ref{eq:renyientropy}) by replacing the  action $S$ by 
 $S_{0}\left[\phi_{+},\phi_{-}\right]$ (i.e. setting $\HB=0$).

The expectation value $\left\langle \cdot \right\rangle _{\RN}$ is taken in the theory living on the Riemann surface $\RN$:
\begin{equation}
\left\langle X\right\rangle _{\RN}=\int_{\RN}{\cal D}\phi_{+}\: X\: e^{-S_{0}\left[\phi_{+}\right]}/\int_{\RN}{\cal D}\phi_{+}e^{-S_{0}\left[\phi_{+}\right]}\label{eq:EVonRN}
\end{equation}
where we have factorized the odd modes  since the interaction $\HB$
does not depend on $\phi_{-}$. \\

Our computation is then a standard perturbative expansion, but it
is carried on a complicated worldsheet. There are  three factors
in Eq. (\ref{eq:perturbDvt}): 
\begin{itemize}
\item 
The quantity $\RNIR$ is the  Renyi entropy at the conformal point. It has been computed in \cite{calabresecardy09}
and we will come back to it in the next section and in the appendix \ref{twistfields}.
\begin{eqnarray}
\RNIR & = & c_{N}\left(\frac{2L}{a_{0}}\right)^{-\frac{c}{6}(N-\frac{1}{N})}\label{eq:RN0}\\
\SIR&=&-\lim_{N\to1}\p_{N}\left(\RNIR\right)=\frac{c}{3}\ln\left(\frac{2L}{a_{0}}\right)+c_{1}^{\prime}
\end{eqnarray}
with $c=1$ the central charge and $c_N$ and $c_{1}^{\prime}$ are non universal constants.
\item  The denominator in (\ref{eq:perturbDvt}) is nothing but the impurity partition function of the system raised to the power $N$, where the impurity partition function is defined as
the ratio of the partition function of the full model $\Cal Z = \tr e^{-\beta(H+\HB)}$ by that of the system as the IR fixed point:
\be
\Cal Z\eimp\equiv\frac{\Cal Z}{\Cal Z^{(0)}}=\vm{e^{-S_{\textsc{b}}}}_{\Cal R_1}
\ee
Its computation involves a series of expectation
values on a single-sheet geometry. For example, in the case of infinite size and zero temperature,
the expectation values are taken on the complex plane and we will see shortly that the denominator
is exactly equal to one.
\item  Similarly, the numerator can be written as a ration of partition functions, but for the system living on a $N$-sheet geometry:
\be
\Cal Z_N\eimp\equiv\frac{\Cal Z_N}{\Cal Z_N^{(0)}}=\vm{e^{-\SB}}_{\Cal R_N}
\ee
We will see in the next section how to compute it perturbatively.
\end{itemize}

\subsubsection{Explicit expansion}

We now turn to the description of the perturbative expansion of the Renyi entropies, or equivalently the partition function on the Riemann surface $\RN$ of the \eemph{full} theory with boundary interaction (\ref{eq:hamIR}). \\
The partition function $\Cal Z_N$ is now an infinite sum of contributions involving 
$k$-uple integrals (see \figref{rhoinsere}) of $k-$points correlators on $\RN$ of operators $\Cal O_{n_i}$, $\vm{\prod_{i=1}^k\Cal O_{n_i}(z_i)}$, where $z_i=\tau_i$ is a point on $\RN$ at position $x=0$. To start with, the exponential $e^{-S_{\textsc{b}}}$ is time-ordered in imaginary time. 

Recall the effective IR hamiltonian is obtained from the expression of the boundary state describing the full interacting theory in the closed-chanel geometry \cite{lesage_perturbation_1999}, whose expression is naturally regularized by inserting the perturbating operators at points $z_j=\tau_j + ix_j$ with $0<-x_1<-x_2<...<-x_k\ll 1$. Note that this regularization of the UV divergences by point splitting is the only one consistent with integrability. Within the point splitting regularization, the $k$-uple integral is done over paths that are slightly displaced and therefore never cross.

Furthermore, recalling that the perturbing operators $\Cal O_n$ are conserved quantities, one  observes  that we can freely exchange the order of two contours. Indeed the term that is generated when two contours $i$ and $j$ are exchanged is simply 
$
...\int dz_j \oint_{z_j} dz_i \Cal O_{n_i}(z_i)\Cal O_{n_j}(z_j) ... =0
$ 
(here $...$ indicates that this expression appears inside some correlators and some integrals). The vanishing of this expression resulting from the $\Cal O_n$'s being commuting conserved quantities, that translates in the OPE: $\Cal O_{n}(z)\cdot\Cal O_m(\omega)\sim ... + \frac{1}{z-\omega}\p \Cal O_{n,m}(\omega)+\mbox{reg.}$ with $O_{n,m}$ a local operator.

We are now in a position to write down the expansion of the impurity Renyi entropy $\frac{R_N}{\RNIR}$:
\bea
R_N\eimp\equiv \frac{R_N}{\RNIR} &=&\sum_{\boldsymbol n} C_{\{ n_i\}}\; A^{\{ n_i\}}
\label{devRNgen}
\\
A^{\{n_{i}\}}&=&\prod_{i=1}^k\int_{\left\{ x=0\right\} }dz_{i}\left\langle \prod_{i}{\cal O}_{2n_{i}+2}(z_{i})\right\rangle _{\RN}
\label{defAn}\\
C^{\{ n_i\}}&=& (-1)^k\,\TB^{-\sum_{i=1}^k (2n_i+1)}\,\prod_{i=1}^p\frac{(g_{2m_i+2})^{q_i}}{q_i !}
\label{defCn}
\eea
where the sum in (\ref{devRNgen}) is taken over all integers vectors $\boldsymbol n=(0\leq  n_1\leq ....\leq n_k)$, and one has introduced the integers $m_i,q_i$ writing 
$\boldsymbol n = (
\underbrace{m_1,...,m_1}_{q_1},\underbrace{m_2,...,m_2}_{q_2},...\underbrace{m_p,...,m_p}_{q_p})
$ with $m_i<m_{i+1}$.
For fixed $n_i$, the integral over
$z_i$ is to be taken along the $N$ possible disconnected paths on $\RN$ going
from $-\infty$ to $+\infty$ in imaginary time, and position $x=0$ (thick line in \figref{RNgeom}).

In the next section we will see how we can compute parts of the $A^{\{n_{i}\}}$  in the case of infinite
size and $T=0$, where the sheets of $\RN$ are the complex plane.

\section{Infinite Size}
\label{infinitesize}

In this section we report results for the perturbative expansion in the simplest case of a system of infinite size
and at zero temperature. We use an orbifold representation of the expectation
values that appear in (\ref{eq:perturbDvt}) and conformal transformations
to explicitly compute the expansion up to order $\displaystyle\frac{1}{(L\TB)^6}$.
We refer to appendices \ref{twistfields} and \ref{emtInOrbifold} for details about twist fields and orbifold representation.

\subsection{Infrared fixed point}

We start by briefly describing the theory at its IR fixed point (formally obtained by setting $\TB=\infty$) focusing on its equivalence   with a $\mathbb{Z}_N$ orbifold theory.
This method was introduced in Ref. \cite{calabresecardy09} by Calabrese and Cardy
to compute the first term $R_{N}^{(0)}$. It relies on the introduction 
of "twist fields" which are inserted in a path integral on
a more simple geometry and which encode the twisted geometry of
$\RN$.

Comparing (\ref{eq:orbiCorrel}) and (\ref{eq:EVonRN}), Calabrese and Cardy
find that $\int_{\RN}{\cal D}\varphi_{+}e^{-S_{0}}\propto\VM{\Phi(u)\tilde{\Phi}(v)}$, where $\Phi,\tilde{\Phi}$ are the twist fields, that are inserted at points $u=-iL$ and $v=+iL$.
This leads to the well-known result  for the Renyi
entropy of an interval of length $2L$ with the rest of the system
in a free theory
\begin{eqnarray}
\RNIR & = & c_{N}\left(\frac{2L}{a_{0}}\right)^{-\frac{c}{6}(N-\frac{1}{N})}\label{eq:RN0BIS}
\end{eqnarray}
where we recall that we have chiral fields but \eemph{two} copies $\phi_{\pm}$, resulting in the factor $c/6$ in the exponent. When taking the limit $N\to 1$, this results in the by-now well-known expression \cite{holzhey94,calabresecardy04,calabresecardy09} for the entanglement entropy at the strong coupling fixed point:
\be
\SIR=\frac{c}{3}\ln\left(\frac{2L}{a_{0}}\right)+ \mbox{non universal}
\label{eq:S0}
\ee
with $c=1$ in our case.

\subsection{Away from the IR fixed point}

\begin{widetext}
\begin{figure}[H]
\begin{centering}
\includegraphics[width=16cm]{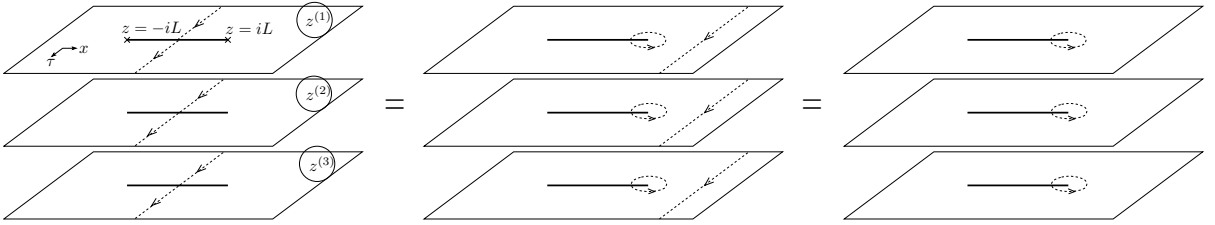}
\par\end{centering}

\caption{\small Deforming one contour on $\RN$, illustrated with $N=3$.}
\label{fig:contours}

\end{figure}
\end{widetext}

A generic term in the perturbative expansion corresponding to the insertion of $k$ boundary operators, consists in a $k$-fold integral on the surface $\RN$. One can actually deform the paths of integration continuously, one by one: the integrands
are $k-$points correlation functions of the ${\cal O}_{i}$'s, therefore
analytical everywhere on $(\RN)^{k}$, except for the branching points
at $z=u,v=-iL,+iL$. Bringing the contours past the singular point $z=v$
leaves behind a contribution that is simply a path $P_{N}$ circling
anti-clockwise the singular point $z=v$ on $\RN$. The other contribution,
being past $z=v$, can be moved to infinity, where it vanishes due to the clustering properties of the $k$-point correlator. This is shown in \figref{contours} in the case $N=3$ and one contour (if there are many contours, one obtains nested integrals around $z=v$).

In fact, since the path of integration $P_N$ is invariant under cyclic permutations of the replicas, we are dealing here only with very special operators, namely $\mathbb{Z}_N$ invariant operators $X^{(0)}(z)=\sum_{j=1}^N X(z^{(j)})=\sum_{j=1}^N X_j(z)$, with $j$ a sheet index.
The equivalence between the theory defined on $\RN$ and an orbifold theory allows to express the mean value of any such operator as:
\begin{equation}
\VM{X^{(0)}(z)}_{\RN}=\VMORB[X^{\iorb}(z)]\label{eq:orbiCorrel2}
\end{equation}
where expectation values in the right hand side are taken in the orbifold theory on the plane. 
Hence, introducing the notation $\res{}{X^{\iorb}\cdot \ttw}=\{X^{\iorb}\cdot \ttw\}_{1}^{\vphantom{\int}}$ for the terms in $\frac{1}{z-w}$ in the OPE $X^{\iorb}(z)\cdot \ttw(\omega)=\sum_{n\in\mathbb{Z}}\frac{1}{(z-w)^n}\,\{X_{\iorb}\cdot \ttw\}_n^{\vphantom{\int}}(\omega)$, one arrives at the point where one can represent, \eemph{at the operator level}, the integral over the path $P_N$:
\be 
\int_{\{x=0\}} dz\vm{ \;\Cal O_n(z)}_{\RN} = 2i\pi\;\frac{\vm{\tw(u)\res{}{\Cal O_n^{\iorb}\cdot\ttw}(v)}}{\vm{\tw(u)\ttw(v)}}
\ee
This is easily iterated (the different contours are nested) with the result:
\begin{eqnarray}
A^{\{n_i\}}(u-v)&=&(2i\pi)^k\;\frac{\vm{\tw(u)\;\res{\vec n}{\ttw}(v)}}{\vm{\tw(u)\,\ttw(v)}}
\\
\res{\vec n}{\ttw}&\equiv&\res{}{\Cal O_{2n_1+2}\cdot...\,\res{}{\Cal O_{2n_k+2}\cdot \ttw}...} \nonumber \\
\label{defResn}
\end{eqnarray}

Considering for a brief moment terms with only $\Cal O_2$ insertions (i.e. $n_1=n_2=...=n_k=0$), and using the fact that $\res{}{T_{\iorb}\cdot X}=\p X$ for any local operator $X$, one immediately finds  
$A^{\{0,...,0\}}(u-v)=\vm{\tw(u)\ttw(v)}^{-1}\,(2i\pi)^k\,\p^k_v \vm{\tw(u)\ttw(v)}$, 
so that 
$\sum_{k\geq 0}\frac{(-g_2/\TB)^k}{k!} A^{\{0...0\}}=\vm{\tw(u)\ttw(v)}^{-1}\vm{\tw(u)\ttw(v-\frac{2i\pi g_2}{\TB})}=\left(1+\frac{1}{L\TB}\right)^{-\frac{c}{12}(N-N^{-1})}$ 
where one has used that $g_2=-\frac{1}{\pi}$, $u=-iL$ and $v=+iL$.
One can thus resum the full series for those terms consisting only on $\Cal O_2$ contributions. 

More generally, since an insertion of $\Cal O_2$ amounts to a derivative with respect to the point $v$, one can resum all $\Cal O_2$ subdiagrams,  $\sum_{k\geq 0}\frac{(-g_2/\TB)^k}{k!} A^{\{n_1,...,n_r,0...0\}}(u-v)=\vm{\tw(u)\ttw(v)}^{-1}\vm{\tw(u)\ttw(v+2i/\TB)}\times A^{\{n_1,...,n_r\}}(u-v-2i/\TB)$:  
This allows for a partial resummation of the perturbative expansion, all $\Cal O_2$ insertions simply amounting to a \eemph{shift} $v\to v'=v+2i/\TB$. We thus  arrive at the final compact form:
\bea
 R_N\eimp&=&\left(1+\frac{1}{L\TB}\right)^{-2h}\; \overline{R}_N\left(u-v-\frac{2 i}{\TB}\right)
\nonumber\\
\Sentang\eimp& = &  \Sentang_{\mathcal{O}_2} +\overline{\Sentang}\left(u-v-\frac{2 i}{\TB}\right) 
\label{Simp:infinite}\\
 \Sentang_{\mathcal{O}_2} & =&\frac{1}{6}\,\ln \left(1+\frac{1}{L\TB}\right) 
\label{eq:resumTN}
\eea
where one has introduced the reduced impurity Renyi entropy free of $\Cal O_2$ insertions:
\bea
\overline{R}_N(u- v') &=& \sum_{\boldsymbol n\;\backslash\;n_i>0} C_{\{ n_i\}}\; A^{\{ n_i\}}(u-v')
\eea
and $\overline{\Sentang}=-\lim_{N\to 1}\p_N \overline{R}_N$. This rewriting considerably reduces the number of diagrams that need to be computed to achieve a given order in the expansion: only diagrams free of $\Cal O_2$ insertions have to be considered, and at the end of the calculation, one replaces  $v'$ by $v'=v+2i/\TB=iL+2i/\TB$.

\subsection{Results}

We now turn to the actual evaluation of the reduced partition function $\overline{\Cal Z}_N(u-v)$. According to Eq. (\ref{defResn}), one way would be to compute nested residues of the perturbing operators $\Cal O_4, \Cal O_6, ...$. The situation is not as simple as for $\Cal O_2$. 

For instance if we want to compute in the
orbifold theory an expectation value such as $\sum_{i}\VMRN[(T^{2})(z^{(i)})]=\sum_{i}\VMRN[(T_{i}^{2})(z)]$
the orbifold version of $\sum_{i}\left(T_{i}^{2}\right)$ is \eemph{not}
$\left(T_{\iorb}^{2}\right)$ and in particular we have a priori no
clue of the OPE of the former with twist fields. 

It is nevertheless possible to compute such an OPE (see appendix \ref{app-O4orb} for the explicit form of 
$\res{}{\Cal O_4^{\iorb}\cdot \ttw}$), but it notably
involves to know in advance the correlators we are at the end interested in. Contrarily to the effect of $\Cal O_2$ insertions, there is (probably)  no simple differential operator representing higher operators $\Cal O_{n}$ at the level of correlation functions, presumably  forbidding a possible resummation of the Renyi entropy in the same spirit as could be done for $\Cal O_2$ contributions. 

The contributions $A^{\{n_i>0\}}$ can nevertheless be evaluated in a straightforward way: any such contribution can be obtained from the knowledge of the $n-$point function of the operator $\Cal O_2$ on $\RN$, by then performing the relevant  normal order operations on the surface $\RN$ (this is the route we have chosen, see for instance appendix \ref{higherorders}). Another equivalent way would be to
derive directly the form of the orbifold operators on $\RN$. For instance, the expression of the operator $\Cal O_4^{\iorb}(z)\eveq\sum_i \mathcal{O}_4(z^{(i)})$ can be deduced from the transformation of the $\Lambda=(T^{2})-\frac{3}{10}\partial^{2}T$
quasi-primary operator, that  reads 

\begin{widetext}
\begin{equation}
\begin{aligned}\Lambda_{\RN}(z) & =f'(z)^{4}\Lambda_{\mathbb{C}}(\omega) +\frac{22+5c}{5c}\cdot\frac{c\{f,z\}}{12}\left(2f'(z)^{2}T_{\mathbb{C}}(\omega)+\frac{c\{f,z\}}{12}\right)
\label{lambdatransformation}
\end{aligned}
\end{equation}
upon the conformal mapping $\omega=f(z),\: z\in\RN,\omega\in\mathbb{C}$ that uniformizes the surface $\RN$ into the complex plane. 

We now turn to the results up to order 7 in $1/(L\TB)$. The partial resummation (\ref{eq:resumTN}) of $\Cal O_2$ insertions leads to a significant simplification of diagrams that need to be computed, from 17 diagrams $A^{\{0\}}$, $A^{\{1\}}$, $A^{\{2\}}$, $A^{\{0,0\}}$, $A^{\{0,1\}}$, $A^{\{0,2\}}$, $A^{\{1,1\}}$, $A^{\{0,0,0\}}$, $A^{\{0,0,1\}}$, $A^{\{0,0,2\}}$, $A^{\{0,1,1\}}$, $A^{\{0,0,0,0\}}$, $A^{\{0,0,0,1\}}$, $A^{\{0,0,0,0,0\}}$, $A^{\{0,0,0,0,1\}}$, $A^{\{0,0,0,0,0,0\}}$, $A^{\{0,0,0,0,0,0,0\}}$ to only 3 reduced diagrams (i.e. diagrams free of $\Cal O_2$ insertions) $A^{\{1\}}$, $A^{\{2\}}$ and $A^{\{1,1\}}$ at order 7.
After tedious calculations, we find:
\bea
\overline{R}_N(u-v) &=& -\frac{g_4}{\TB^3} \; A^{\{1\}}(u-v) - \frac{g_6}{\TB^5} \; A^{\{2\}}(u-v) + \frac{g_4^2}{2\TB^6} \; A^{\{1,1\}}(u-v)  +\Cal O\left((L\TB)^{-8}\right)
\nonumber
\eea
\bea
A^{\{1\}}(u-v) &=& \frac{-i\pi (N^2-1)^2}{24 D\;  N^3(u-v)^3} (4-D)(1-4D)
\nonumber\\
A^{\{2\}}(u-v)&=& \frac{-i\pi \left(N^2-1\right)^2}{5760 D^2 N^6 (u-v)^5}   
\Bigg[24 (2-D) (1-2 D) \left((1-D)^2 (334 N^2-246) - D (64 N^2-31)\right)
\nonumber\\
&&+D N \left(-18 D (1123 N^2-667)+(1-D)^2 \left(7 N^4 11930 N^2 - 5697\right)\right)
\Bigg]
\nonumber\\
A^{\{1,1\}}(u-v)&=& - \frac{\left(
N^2-1\right) \pi ^2}{N^6 (u-v)^6}\;\;
\Bigg[
\frac{(1-4 D)^2 (4-D)^2}{2880 D^2}
 \left(5 (N^2-1)^3+\frac{144}{13} N (-36+79 (N^2-1))\right) \nonumber\\
&&+9 N \left(20-29 N^2+13 N^4\right) \frac{3521 D^2-1576 D \left(1+D^2\right)+216 \left(1+D^4\right) }{1820\; D^2}
\Bigg]
\label{resultsAi}
\eea
Due to the prefactors $(N^2-1)^2$ in all but one of the contributions to $\overline{R}_N$,  the reduced entanglement entropy
bears only one contribution up to order 7, that can be expressed as a function of parameter $\alpha=\frac{(1-D)}{\sqrt{D}}$:
\be
\overline{\Sentang}(u'-v') =  \frac{18}{35}\; \frac{(\pi g_4)^2}{(u'-v')^6\;\TB^6}\;\left(4\alpha^4-8\alpha^2+9\right)
+\Cal O\left((L\TB)^{-8}\right)
\label{resultSinfinite}
\ee
where $u',v'$ are the generic points where the twist fields are inserted in order to compute the quantities $A^{\{n_i\}}$ with $n_i > 0$. Now, taking $u'=-i L, v'=i L + 2i/\TB$ in this expression and keeping only terms that are  $\Cal O\left((L\TB)^{-7}\right)$ leads to the expression for the full entanglement entropy:
\begin{eqnarray}
\Sentang &=& \SIR + \Sentang\eimp\label{result:infinite}\\
{\cal S}\eimp&=&\frac{1}{6}\,\ln \left(1+\frac{1}{L\TB}\right)  - {18\over 35} {(\pi g_4)^2\over (2L\TB)^6}\left(1-\frac{6}{L\TB}\right)(4\alpha^4-8\alpha^2+9)+{\cal O}((L\TB)^{-8})
\label{result:infinite:imp}
\end{eqnarray}
with $\SIR$ the entanglement entropy at the IR fixed point given by (\ref{eq:S0}), and we recall the 
$D-$dependence of the coefficients $g_4$ and $\alpha$:
\begin{equation}
g_4={D\over 6\pi^2} \left({\Gamma(D/2(1-D))\over \Gamma(1/2(1-D))}\right)^3{\Gamma(3/2(1-D))\over \Gamma(3D/2(1-D))},~~\alpha=\frac{(1-D)}{\sqrt{D}}.
\end{equation}
Note that in (\ref{result:infinite:imp}), the first term in the right hand side  has  to be truncated at order 7.

The second term in the right-hand side of equation (\ref{result:infinite:imp}) is the first term in the perturbative expansion that depends on the strength of interactions $U$ in the original model. We note that its sign is negative, and this is consistent with the fact that   
the impurity entanglement entropy should eventually saturate at the value $\ln (2)$ in the limit $L\TB\to 0$. An interesting observation is that this correction is \eemph{minimal} at the free point $U=0$ in the IRLM or the Toulouse point in the Kondo model, see \figref{SfixedL}. 

In the regime of attractive coulombic interaction, i.e. at negative $U$ corresponding to $D>1/2$, the correction (\ref{resultSinfinite}) displays a very weak dependence on $D$, and the entanglement entropy, at order $(L\TB)^{-7}$, is essentially that of free fermions $U=0$.

On the other hand, at very small $D$ (this situation can be reached in the one-wire IRLM by choosing a sufficiently large repulsive coulombic interaction $U$), the correction  (\ref{resultSinfinite}) scales as $1/D^4$ 
and  is thus unbounded: the universal crossover function defining the entanglement entropy is thus deeply affected by the interactions for $U>0$, even in the strong coupling regime.

\begin{figure}[h]
\begin{centering}
\includegraphics[width=10cm]{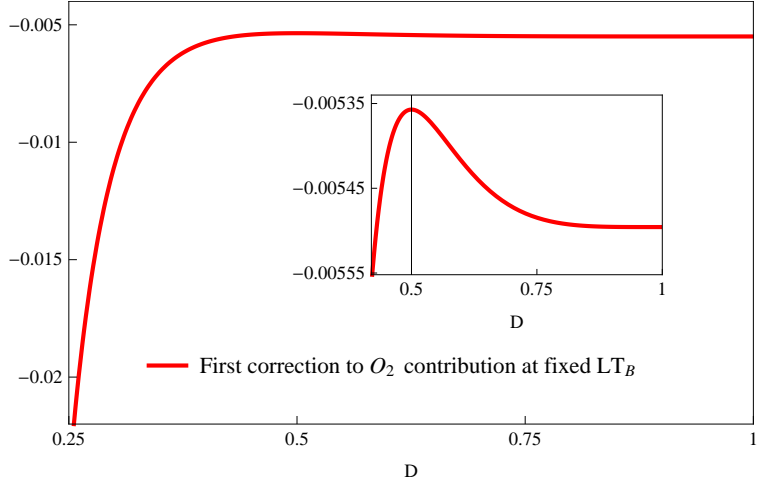}
\par\end{centering}

\caption{\small Impurity entanglement entropy: first correction to the contribution $\frac{1}{6}\ln (1+1/L\TB)$ stemming from $\Cal O_2$, displayed here at fixed $L\TB=1$ as a function of the interactions. $D=\frac{1}{2}$ corresponds to the free point, where this correction is minimal (see inset).  A significant enhancement of this negative correction is visible when reducing $D$.}
\label{fig:SfixedL}

\end{figure}
\end{widetext} 

It is instructive to compare the universal crossover function for the entanglement entropy with that of the boundary entropy $s$ introduced by Affleck and Ludwig \cite{affleckludwig1,affleckludwig2}. In the UV limit, both those quantities reach the value $\ln (2)$. In the IR limit, they both vanish. The boundary entropy $s$ is simply related to the partition functions of the system at finite temperature $\beta^{-1}$: $s=(1-\beta\p_\beta)\ln  \Cal Z\eimp$
with $\Cal Z\eimp=\frac{\Cal Z(\beta)}{\Cal Z^{(0)}(\beta)}$ and $\Cal Z^{(0)}$ the partition function of the system without impurity. One immediately gets $\ln  \Cal Z\eimp=\ln \vm{e^{-\SB}}_\beta$ where the expectation value is taken at temperature $\beta^{-1}$, i.e. on a cylinder of circumference $\beta$. Because the perturbing operators $\Cal O_n$ are commuting conserved quantities, in the perturbative expansion of this quantity around the strong coupling fixed point, the contours can be deformed and moved one by one to infinity along the cylinder, resulting in $\vm{e^{-\SB}}_\beta=e^{-\vm{\SB}_\beta}$ and finally $\ln  \Cal Z\eimp=-\vm{\SB}_\beta=-\beta\sum_n g_{2n+2}\TB^{-(2n+1)}\vm{\Cal O_{2n+2}}_\beta$. 
The boundary entropy thus has an expansion in odd powers of $T/\TB$. 
Evaluating the one-point functions $\vm{\Cal O_{2n+2}}_\beta$ in the cylinder geometry with circumference $\beta$, one arrives at the expansion
\be
s = -\pi g_2 \frac{1}{3} \left(\frac{\pi}{\beta\TB}\right)
-\pi g_4 \frac{9 -4 \alpha^2}{15} \left(\frac{\pi}{\beta\TB}\right)^3
-\pi g_6 \frac{96\alpha^4-340\alpha^2+425}{252} \left(\frac{\pi }{\beta\TB}\right)^5+\Cal O\left((1/\beta\TB)^7\right).
\ee
We note that a dependence in the  interaction parameter $D$ appears at order 3 already, making the boundary entropy more sensitive to interactions than the entanglement entropy.

\section{Finite Size}
\label{finitesize}

We now want to compute perturbatively the expression of the entanglement
entropy around the IR fixed point but in the case when the size of
the system is finite, of length $L_0$ in the original geometry or $2L_0$ in the unfolded geometry. In this case, the replica trick leads to expressions
of Renyi entropy similar to Eq. (\ref{eq:renyientropy}), but instead of considering
a surface $\RN$ made with $N$ infinite planes with branching cuts,
one gets a surface $\CN$ made of $N$ cylinders periodic along the $x$ direction 
 with the same branching cuts between them. 

\begin{figure}[h]
\begin{centering}
\includegraphics[width=6cm]{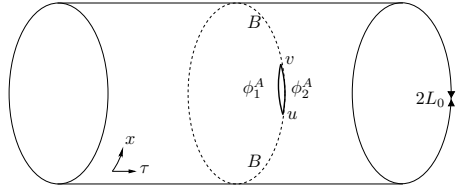}
\par\end{centering}

\caption{\small The reduced density matrix is now a cylinder, periodic in space, with a cut along the $x$ axis. The $\CN$
geometry  now consists in $N$ cylinders connected along the cuts. The partial trace over subsystem $B$ results in the gluing of two semi-infinite cylinders along the dashed line, at $\tau=0$. The impurity is inserted at $x=0$ (not represented).}
\label{fig:cylindre_taille_finie}

\end{figure}

\subsection{Infrared fixed point and twist fields}

The first contribution to the entanglement entropy at finite size in the strong coupling limit $L\TB\to \infty$ is, as before, the entanglement entropy at the IR fixed point.  This quantity, that was computed by  Calabrese and Cardy \cite{calabresecardy09}, can be obtained as the two-point function of the twist and anti-twist operator at finite size  ; since those operators are primary operators, by using the conformal mapping that sends the cylinder onto the plane:
\be
z\in \Cal C\too \omega = e^{\frac{\pi}{L_0}z} \in \mathbb{C}
\label{CylToPlane}
\ee
one easily computes 
 $\left\langle \Phi(u)\tilde{\Phi}(v)\right\rangle _{L_{0}}$, yielding:
\begin{equation}
\SIR_{L_0}=\frac{c}{3}\ln \left(\frac{L_{0}}{\pi a_{0}}\sin\frac{\pi L}{L_{0}}\right)+c_{1}^{\prime}.
\label{S0finitesize}
\end{equation}
Here $a_{0}$ is the inverse of the ultraviolet cut-off (e.g. in a lattice regularization, it coincides with the lattice constant), and $c_{1}^{\prime}$ is again a non
universal constant.

The finite size infrared fixed point is now our reference point around which one wishes to expand the Renyi and entanglement entropies.
The orbifold representation allows one to compute the expansion even
when the surface is $\CN$. Indeed, since this representation gives
the correlators of the original $N$-sheeted geometry in terms of correlators
of the orbifold theory in the geometry of a lone sheet, one only has  to perform the 
change of variables from the cylinder to the complex plane (\ref{CylToPlane}). 

\begin{figure}
\begin{centering}
\includegraphics[width=7cm]{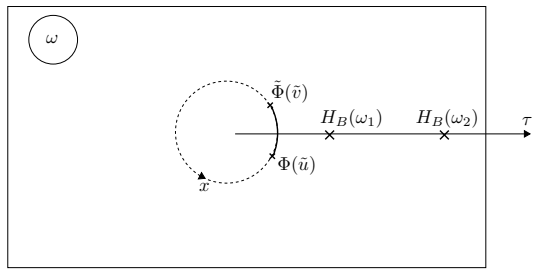}
\par\end{centering}

\caption{\small 
The  orbifold theory is now living on a cylinder and one has to uniformize
it to the complex plane. The integration contour corresponding to the impurity insertion is then radial and
goes from $\omega=0$ to $\omega=\infty$, passing between the twist
and anti-twist fields. The dashed line corresponds to the partial trace over subsystem $B$.}
\label{fig:finite-size-integration}

\end{figure}

However, there is a slight difference with before. When considering
the surface $\RN$, one can move the integration contours freely in
the orbifold representation and only get a contribution from residues
around one of the $u,v$ points, and this fact greatly simplifies the computations.
In the finite size case,  we first have to flatten the
cylinder and change the integration contour accordingly (see Fig. \ref{fig:finite-size-integration}). In the new geometry the integration contour on the plane is not a
circle anymore, it is instead not  not closed leading to more
complicated integrals. It results that the insertion of the operator ${\cal O}_{2}$
has no simple action and we cannot derive an equality such as Eq. (\ref{eq:resumTN}).

\subsection{IR divergences}

At finite size,  one also has  to be careful to take into account the expansion
stemming from the impurity partition function $\Zimp=\left<e^{-\SB} \right>_{\mathcal{R}_1}$ (the denominator in Eq. (\ref{eq:renyientropy})), which is now not equal to $1$ because of the non trivial expectation value of $\Cal O_2$, and of higher order perturbing operators, on a cylinder.
 
This additional complication is easily understood by just considering the first order perturbing operator $\Cal O_2$: having put the system on a cylinder, the curvature induces a non-vanishing local energy density, $\vm{\Cal O_2}=\vm{T}=e_0=-\frac{c\pi^2}{6 L_0^2}$. It results that the impurity partition function is now formally infinite : introducing a (small) finite temperature $\beta^{-1}$ for a moment to regularize those divergences, one has $\Zimp=\lim_{\beta\to\infty} e^{-e_0\frac{\beta}{L_0}}$, so that each term of the expansion of $\Zimp$ in inverse powers of $\TB$  is formally infinite at vanishing temperature.

These divergences will also appear in the $N-$sheeted geometry $\RN$, when computing $\Zimp_N$.
The impurity partition functions $\Zimp$ and $\Zimp_N$ cannot be defined on their own and the quantities $A^{\{n_i\}}$ defined in (\ref{defAn}) as constituents of $R_N$ are now ill-defined.

In fact these divergences compensate term by term in the expansion in $1/L\TB$ of $R_N/\RNIR$:  the ratio between the impurity partition functions is perfectly well defined yielding  a finite answer for the Renyi and entanglement entropies.
One way to see this is to take the logarithm of the Renyi entropy  :
\begin{align}
\ln R_N - \ln \RNIR =& \ln \vm{e^{-S_B}}_{\CN} - N \ln \vm{e^{-S_B}}_{\Cal C_1} = \vm{e^{-S_B}}^c_{\CN} - 1 - N (\vm{e^{-S_B}}_{\Cal C_1}^c -1 )
\end{align}
where we used the well-known relationship between the generating functionals of connected and disconnected expectation values of operators. The connected expectation value of a product of operators $\Cal O_{n_i}(z_i)$ is defined recursively as 
\begin{align}
\vm{\Cal O_n(z)}^c_{\CN} &= \vm{\Cal O_n(z)}_{\CN}
\end{align}
and
\begin{align}
\vm{\Cal O_{n_1}(z_1)\dots\Cal O_{n_k}(z_k)}_{\CN}&=\vm{\Cal O_{n_1}(z_1)\dots\Cal O_{n_k}(z_k)}^c_{\CN} + \sum_{part.} \VM{\prod_{i\in\alpha}\Cal O_{n_i}(z_i)}_{\CN}^c\VM{\prod_{j\in\beta}\Cal O_{n_j}(z_j)}_{\CN}^c \dots
\end{align}
where $\alpha,\beta,\dots$ are non trivial subdivisions of a partition of $[1,k]$. 

This writing allows to regroup the terms of the expansion in $1/\TB$ in well defined quantities.
Indeed, these connected  (or non factorizable from the point of view of the integration variables) contributions of $\VM{\Cal O_{n_1}^{orb}(z_1)\dots \Phi(u)\tilde \Phi(v)}_{\Cal C}/\VM{\Phi(u)\tilde \Phi(v)}_{\Cal C}$ can be written as the sum of two terms.

The first term  depends on $u, v$ has to vanish when $u \to v$ (or $u,v \to \mp iL_0$, cutting open the cylinders) and  encodes  the local influence of the insertion of twist fields at $u, v$. This first contribution does not produce infinities when sending all the $z_i\to \infty$ (while keeping $u$ and $v$ finite) because of the clustering properties of the connected correlator between the $\Cal O_{n_i}$ and $\Phi,\tilde\Phi$.

The second contribution which does not depend on the points $ u,  v$ (from $\VM{\Cal O_{n_1}^{orb}(z_1)\dots}^c_{\Cal C}$) can be shown to be divergent when \emph{all} the points $z_i$ are sent to infinity. In this limit, the fields inserted at points $z_i$ do not feel any longer the cuts on the cylinders, and the correlator is asymptotically that of fields living on $N$ \emph{disconnected} cylinders, yielding a factor $N$, the number of copies, for the result of the integral.

In a nutshell, it can be shown that $\VM{(-S_B)^k}_{\CN}^c= (u - v) F_1(u, v) + N \times\textit{diverging part}$, where the diverging part is independent of $N$ and is exactly compensated by  $N\VM{(-S_B)^k}_{\Cal C_1}$. This shows that the expansion in  $\ln R_N - \ln \RNIR$ are well defined.

As a result, when  computing the Renyi entropy $R_N$ at finite size,  one only has  to take into account the connected correlators in $\Cal Z\eimp_N$, and do not care about the diverging parts that are in fact all canceled out by the denominator $\Cal Z\eimp$ at every order in the development in $1/\TB$.

\subsection{Twist fields at finite size}

We can now easily  generalize the calculation of the first order terms to the finite size case. 
Under the conformal mapping (\ref{CylToPlane}), the singular points $u,v$ of the are mapped onto $\tilde v=e^{i\pi L/L_0}$ and $\tilde u=e^{-i\pi L/L_0}$, so that one has for instance:

\begin{align}
\sum_{i=1}^{N}\left\langle T(z^{(i)})\right\rangle _{\CN} & =\frac{\VMsurface[{\cal C}]{T_{\iorb}(z)\Phi(u)\tilde{\Phi}(v)}}{\VMsurface[{\cal C}]{\Phi(u)\tilde{\Phi}(v)}}\\
 & =\frac{\VMsurface[\mathbb{C}]{\left(\frac{dz}{d\omega}\right)^{-2}T_{\iorb}(\omega)\Phi(\tilde u)\tilde{\Phi}(\tilde  v)}}{\VMsurface[\mathbb{C}]{\Phi(\tilde u)\tilde{\Phi}(\tilde v)}}\nonumber -\frac{\hat{c}}{12}\left(\frac{dz}{d\omega}\right)^{-2}\left\{ z;\omega\right\} 
\end{align}

The last term yields a diverging contribution that is exactly canceled out by a term coming from $(\Zimp)^N$, the denominator of Eq. (\ref{eq:renyientropy}), so that we get:
\begin{equation}
A_{L_{0}}^{\{0\}} =\frac{\pi }{L_{0}}\int_{0}^{\infty}d\omega
\left[
\frac{h\omega(\tilde u-\tilde v)^{2}}{\left(\omega-\tilde u\right)^{2}\left(\omega-\tilde v\right)^{2}} -\frac{N}{24 \omega}\right]
=-\frac{2\pi h}{L}f_{1}(L/L_{0})+div.  
\end{equation}
with 
 the function $f_{1}(a) =a\left(1+\pi(1-a)\cot\pi a\right) $. Similarly one can obtain the higher order contributions involving insertions of the perturbing operator $\Cal O_2$ (see Appendix \ref{twist@finite}).

The next step consists in computing correlators of higher order perturbing operators $\Cal O_{n>2}$, on the surface $\CN$. One route would be to express the $\mathbb{Z}_N$-neutral combinations corresponding to those operators, namely $\Cal O_n^{\iorb} = \sum_{j=1}^N\Cal O_n\left(z^{(j)}\right)$, in terms of orbifold (quasi-) primary operators. We have chosen another equivalent route, namely
a quantization method that allows to directly compute
correlators of more complicated operators like $\mathcal{O}_4$.

\subsection{Direct quantization}

A direct way to compute the correlators $A^{\{n_{i}\}}$ at finite
size is to look directly at the structure of the Hilbert space
in the twisted sector and to consider a quantized version of the fields.
This is a generalization of the quantization of fields with Neveu-Schwarz
or Ramond conditions. The field living on $N$ sheets is in fact a
$N-$valuated field and we shall quantized it this way.

\subsubsection{Monodromy relations}

To be general, in the replica trick we consider a CFT on a Riemann
surface with $N$ sheets. Each sheet is described by the complex coordinate
$z=\tau-ix$. We consider a real free scalar field $\varphi(z,\bar{z})$
living on the Riemann surface, its value on a given sheet $j=1,\dots,N$
is denoted by $\varphi_{(j)}(z)$, so that we have in fact
$N$ copies of a (chiral) CFT. 

These copies are not independent : we consider a branch point at $z=0$,
such that $\varphi_{(j)}(0^{+}-ix)=\varphi_{(j-1)}(0^{-}-ix)$, for
all $x>0.$ In other words, when one turns anticlockwise around the
point $z=0$, one decreases the index of the sheet by one :
\begin{equation}
\varphi_{(j)}(ze^{2i\pi})=\varphi_{(j-1)}(z)
\end{equation}
making afterwards the conformal transformation (\ref{eq:transfoplanplan})
will allow one to consider the branch cut that we are interested in.
It is more convenient to consider the Fourier-conjugate fields, in order
to diagonalize the monodromy relations. We thus introduce the fields
$\varphi^{(k)}(z,\bar{z})=\frac{1}{\sqrt{N}}\sum_{j=1}^{N}e^{2i\pi jk/N}\varphi_{(j)}(z,\bar{z})$,
which verify the relations
\begin{equation}
\varphi^{(k)}(ze^{2i\pi})=e^{\frac{2i\pi k}{N}}\varphi^{(k)}(z),\label{eq:boundarycond}
\end{equation}
and since the holomorphic energy-momentum tensor is given by $T(z)=- 2\pi \sum_{i}:\partial\varphi_{(i)}\partial\varphi_{(i)}:(z)$,
we also have
\begin{equation}
T(z)= -2\pi \sum_{k}:\partial\varphi^{(-k)}\partial\varphi^{(k)}:(z)=\sum_{k}T^{(k)}(z).\label{eq:decompositionEMT}
\end{equation}
Note that $\varphi^{(-k)}\equiv\varphi^{(N-k)}=\varphi^{(k)*}$ : the
index is to be understood modulo $N$.
\footnote{
If $N$ is even, we now have a free theory with two real bosonic fields
($\varphi_{(0)},\varphi_{(N/2)}$ and the  associated energy-momentum tensor 
is $T^{(0)},T^{(N/2)}$) and $(N-2)/2$ complex bosonic fields (the associated energy-momentum
tensors are the $T^{(k)}+T^{(-k)}=-4\pi(\p\varphi^{(k)}\p\varphi^{(-k)})$).
One the other hand,  if $N$ is odd, we have only one real field $\varphi_{(0)}$
and $(N-1)/2$ complex fields.}

\begin{figure}[h]
\begin{centering}
\includegraphics[width=6cm]{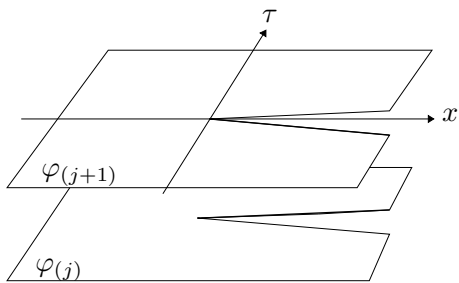}
\par\end{centering}

\caption{\small
Fields living on a $N-$sheeted Riemann surface. The cut is along
the $x$ axis, between $u=0,v=\infty$.}

\end{figure}

\subsubsection{Mode decomposition}

A canonical quantization on a cylinder altogether with imposing the
boundary conditions (\ref{eq:boundarycond}), leads to the mode decomposition
\begin{align}
\varphi^{(k)}(z) & =\frac{i}{\sqrt{4\pi}}\sum_{m}\frac{1}{m}a_{m}^{(k)}z^{-m}\qquad m\in\left(\mathbb{Z}-k/N\right)
\label{modedecomposition}
\end{align}
with the commutation relations 
\begin{equation}
\left[a_{m}^{(k)},a_{n}^{(k^{\prime})}\right]=m\delta_{k+k^{\prime}}\delta_{m+n}\label{eq:twisted algebra}
\end{equation}

Acting on the \emph{twisted} vacuum we have $a_{m}^{(k)}\twistedvac=0$
if $m>0$. Here $\twistedvac=\bigotimes_{k=0}^{N-1}\ktwistedvac$
is the twisted vacuum and the $\ktwistedvac$ are the highest-weight
vectors of the representation of the algebra (\ref{eq:twisted algebra}) ;
$k$ is the index of the monodromy sector. 

The twist field $\Phi(z)$ is the primary field associated with $\twistedvac$,
and it can be written as a product of the twist fields associated
with $\ktwistedvac$ : $\Phi(z)=\Pi_{k}^{\prime}\Phi^{(k)}(z)$. The
prime on the product simply denotes that we count $k$ only for the
independent fields. We shall see in the next paragraph that the conformal
dimension of $\Phi_{k}$ is $h_{k}=(2)\frac{k(N-k)}{4N^{2}}$, the
factor 2 depending on whether $k$ is the index of a real or complex
field. In the twisted sector the vacuum expectation values are abbreviated
as $\VM{X}\equiv\frac{\left\langle \sigma\right|X\left|\sigma\right\rangle }{\left\langle \sigma\right|\left|\sigma\right\rangle }$.

Within this canonical quantization scheme, one can build the perturbing operators $\Cal O_n$ and evaluated arbitrary correlators thereof by using Wick's theorem for the bosons $\varphi^{(k)}$.  Finite size is eventually taken into account by means of the conformal  mapping (\ref{CylToPlane}).
We report on the details of the actual computations in Appendix \ref{directquantization} and directly turn to the end result for the entanglement entropy.

\subsection{Results}

Finally let us summarize the results for the entanglement entropy
of a segment of width $2L$ in a system at finite size $2L_{0}$,
up to order ${\cal O}(\TB^{-4})$.
The net expression for the entanglement entropy is:
\begin{eqnarray}
\Sentang & =&\SIR_{L_0}+\Sentang\eimp\label{result:finite}\\
\Sentang\eimp &=&\frac{1}{6L\TB}f_{1}(L/L_{0})-\frac{1}{12(L\TB)^{2}}f_{2}(L/L_{0})\nonumber\\
& & +\left(\frac{1}{2L\TB}\right)^{3}\left( 
\frac{4}{9}f_3(L/L_0) + \frac{(2\pi)^3\,g_4}{120}(9-4\alpha^2)(L/L_0)^2 f_1(L/L_0) 
\right)
\label{result:finite:imp}
\end{eqnarray}
where $\SIR_{L_0}$ is the Calabrese-Cardy entanglement entropy (\ref{S0finitesize}) in the IR limit $L\TB\to\infty$, and the impurity entanglement entropy is expressed in terms of the functions (we also give  their asymptotics when $a=L/L_0\to 0$):
\begin{align}
f_{1}(a) & =a(1+\pi(1-a)\cot a\pi)\sim1 - \pi^{2}a^{2}/3 +\pi^2 a^3/3+\mathcal{O}(a^4)\\
f_{2}(a) & =f_1(a)^2+a^2(1-a)^2 \pi^2 \sim 1 + \pi^2 a^2/3 - 4\pi^2 a^3/3 +\mathcal{O}(a^4)\\
f_{3}(a) & =f_1(a) f_2(a) + 2 a^3(1-a)^2 \pi^2 \sim 1 +\pi^2 a^3 +\mathcal{O}(a^4) 
\end{align}
We can see that  the interactions (which are hidden in
the coefficients $\alpha=\frac{1-D}{\sqrt{D}}$ and $g_4$) 
do not affect  the entanglement entropy before the
third order in $1/L\TB$: the effect of the interactions at strong coupling
is thus a subtle one, just as in the infinite size case. We again note that this conclusion is invalidated at strong repulsive Coulomb interaction in the one-wire IRLM, where $D$ can be made as small as wished: in this limit, 
the term proportional to $g_4$ scales as $1/D^2$ and thus eventually becomes dominant at sufficiently strong $U>0$.

Our result for the entanglement entropy has the expected property that $\Sentang\eimp$ vanishes order by order in the limit $L\to L_0$: in this limit, region B shrinks to a point (and this limit is not singular since region B does not contain the impurity) so that all Renyi entropies evaluate to 1.

\section{Conclusions}

The entanglement entropy is an elusive quantity, which in many respects differs from usual local observables. 
In 1+1 dimensional many-body systems at their critical point described by a CFT, it has been shown that the replica method allows for a computation of the Renyi and the entanglement entropies.
In this paper, we have shown how to combine the replica method with infrared perturbation theory to compute the bipartite entanglement entropy of a segment of length $2L$ with the rest of the system in the case of a CFT perturbed at its boundary by the interaction with an impurity, namely the Interacting Resonant Level Model, or the one-channel anisotropic $s=1/2$ Kondo model. Our approach allows for a determination of the Renyi and entanglement entropies in a systematic expansion around the strong coupling fixed point that corresponds to the limit $L\gg\TB$, where $\TB$ is the boundary energy scale generated by the interaction with the impurity. We have also investigated the case of a finite size system, and shown that a similar expansion can be carried out in this case. 

Our results indicate that generically, the effect of the interactions on the entanglement entropy is weak: the first contribution to the entanglement entropy that depends on the interactions  scale as $1/(L\TB)^6$ at infinite size, and as $1/(L_0^2 L \TB^3)$ at finite size. On the other hand, we emphasize that the effect of interactions is strongly enhanced in the case of strong coulombic repulsion $U>0$ in the one-wire IRLM. 

Note that our method could also be extended to more complicated geometries, like finite temperature \eemph{and} finite size, the only (but significant regarding the difficulty in computing the correlators!) difference being that  the orbifold theory should be put on a torus.

At the end of  such a technical paper, it is of course natural to ask how the perturbative expansion compares with say a numerical evaluation of the quantity $\Sentang\eimp$ over the whole flow from UV to IR, and how, in particular, the UV limit - which is non perturbative in $\gamma$  - is approached. This will be discussed elsewhere \cite{BFSVP}. Similarly, it is clear that in most circumstances, the quantity of physical interest will not be be $\Sentang\eimp$ itself, but rather the entanglement between the two wires connected by the resonant level. This will involve more technicalities, and will also be discussed elsewhere. 

\bigskip
\noindent{Acknowledgments}:  We thank Ian Affleck, John Cardy, Pasquale Calabrese and Peter Schmitteckert for discussions. HS was supported by the French Agence Nationale pour la Recherche (ANR Projet 2010 Blanc SIMI 4 : DIME) and the US Department of Energy (grant number DE-FG03-01ER45908). LF acknowledges the support of the French Direction G\'en\'erale pour l'Armement.

\appendix

\section{Twist fields and orbifold representation}
\label{twistfields}

We consider a CFT of central charge $c$ (later one will apply this to the case $c=1$) with fields living on the
surface $\RN$ with branching conditions between the $N$ sheets of the surface.
As argued in Refs. [\onlinecite{dixon87,bershadsky88}], 
this is equivalent to considering a theory leaving on the complex plane  but with
twisted vacuums $\lorbivac=\Phi(u)\vac$ and $\rorbivac=\Rvac\tilde{\Phi}(v)$
and central charge $\hat{c}=Nc$. The equivalence as the following meaning: given a local operator $X(z^{(j)})$ in the original theory defined on $\RN$ (here $z$ belongs to the punctured complex plane $\mathbb{C}\backslash\{u,v\}$ and $j=1...N$ is a sheet index), one can form linear combinations $X^{(k)}(z)=\sum_j e^{2i\pi jk/N} X(z^{(j)})$ that carry a well defined charge under the $\mathbb{Z}_N$ cyclic exchange  of the sheets. The equivalence states that to each field $X^{(k)}$ there corresponds a field $X_{\iorb}^{(k)}$ in the twisted theory such that the correlators of $X^{(k)}$ fields on $\RN$ coincide with those of $X_{\iorb}^{(k)}$ according to:
\be
\vm{...X^{(k)}(z)...}_{\RN} = \VMORB[...X_{\iorb}^{(k)}(z)...]
\label{eq:orbiCorrel}
\ee
and in the following we will denote $X^{(k)}\eveq X_{\iorb}^{(k)}$ this equivalence between operators.

The twist (and anti-twist) fields $\Phi$ (and $\tilde{\Phi}$),
are primary fields of conformal dimension $h=\frac{\hat{c}}{24}\frac{(N^{2}-1)}{N^{2}}$. 
(Note that in this paper the CFT one considers are chiral ones as the result of our unfolding of the original problem, so that $\bar h$ is not defined).
$\vac$ is the vacuum of the chiral theory living on a single sheet. The
two point function on the plane of these twist fields reads : 
\begin{equation}
\left\langle \tw(u) \ttw(v)\right\rangle =1/\left(u-v\right){}^{2h}
\end{equation}

\section{Energy momentum tensor in the orbifold theory}
\label{emtInOrbifold}

In this appendix one derives some identities for the operator product expansion of the energy-momentum tensor and its "orbifold powers", by using the equivalence between operators in the untwisted sector of the orbifold theory and $\mathbb{Z}_N$-neutral combinations of operators in the original theory living on the Riemann surface $\RN$.

\subsection{The orbifold energy-momentum tensor at infinite size}
\label{app-Torb}

We start by considering the $\mathbb{Z}_N$-neutral combination $T^{(0)}(z)=\sum_{j=1}^N T_{\RN}(z^{(j)})=\sum_{j=1}^N T_j(z)$
of the stress energy tensor (here $j$ labels the sheets of the Riemann surface $\RN$), and show, following Ref. \onlinecite{calabresecardy09}, that it coincides with the orbifold energy-momentum tensor $T_{\iorb}$ (and we will write $T_{\iorb}\eveq\sum_j T_j$) in the sense that 
\begin{equation} 
\vm{X}_{\RN} = \frac{\vm{X_{\iorb}\,\tw(u)\ttw(v)}}{\vm{\tw(u)\ttw(v)}}
\label{genEquiv}
\end{equation} 
where $X=T^{(0)}(z_1)...T^{(0)}(z_k)$ is a collection of energy-momentum tensors inserted at arbitrary points on $\mathbb{C}\backslash\{u,v\}$. The left hand side of Eq. (\ref{genEquiv}) (first route) can be computed by using the uniformizing map
that sends $\RN$ onto the punctured  complex plane:
\begin{equation} 
z\too\xi=\frac{z-v}{z-u}\too\omega=\xi^{1/N}.
\label{tfo}
\end{equation} 
On the other hand (second route), the right-hand side of Eq. (\ref{genEquiv}) can be computed using a Ward identity involving the energy-momentum tensor $T_{\iorb}$.

\subsubsection{One point correlator}

Let us follow the first route to compute the one-point function of $T^{(0)}$ on $\RN$. Under the uniformizing transformation (\ref{tfo}), the energy-momentum tensor transforms as:
\begin{equation} 
T_{\RN}(z)=(\p_z\omega)^2\left[ T_{\mathbb{C}}(\omega)+\frac{\kappa}{\omega^2}  \right]\qquad\qquad \kappa=\frac{c}{24}(N^2-1)
\end{equation} 
so that 
\begin{equation} 
\vm{T_{\RN}(z)} = \kappa\left(\frac{\p_z\omega}{\omega}\right)^2=\frac{\kappa}{[N(v-u)]^2}\frac{(1-\xi)^4}{\xi^2}
\end{equation} 
where one used 
$\p_z\omega = \frac{\omega}{N(v-u)}\,\frac{(1-\xi)^2}{\xi}
$.
Since this one-point function doesn't depend on the Riemann subsheet where it is evaluated, one thus gets the final result by just multiplying by $N$:
\begin{equation} 
\vm{T^{(0)}(z)}_{\RN}=\frac{\kappa(1-\xi)^4}{N(u-v)^2\xi^2} = \frac{\kappa}{N}\,\frac{(u-v)^2}{(z-u)^2(z-v)^2}.
\label{<T>}
\end{equation} 

According to the second route, one also has: $\vm{T^{(0)}(z)}_\RN=\frac{\vm{T_{\iorb}\tw(u)\ttw(v)}}{\vm{\tw(u)\ttw(v)}}$ with $T_{\iorb}$ the energy-momentum tensor of the orbifold theory with central charge $\hat c=cN$, and $\vm{\tw(u)\ttw(v)}=\left(u-v\right)^{-2h}$ where $h$ is the conformal dimension of the twist fields.  
As noted in Ref. \onlinecite{calabresecardy09}, one can indeed recover the result (\ref{<T>})
by using the fundamental OPE:
\begin{equation} 
T_{\iorb}(z)\cdot\tw(u) = \frac{h\tw(u)}{(z-u)^2}+\frac{\p_u\tw(u)}{z-u}+\mbox{reg.}
\label{OPETtau}
\end{equation} 
that leads to the following Ward identity:
\begin{equation} 
\vm{T_{\iorb}(z)\tw(u)\ttw(v)} = \left[
\textstyle \frac{h}{(z-u)^2} + \frac{h}{(z-v)^2} +
\frac{\p_u}{z-u}+\frac{\p_v}{z-v} \right]\;\vm{\tw(u)\ttw(v)}
= \frac{h(u-v)^{2-2h}}{(z-u)^2(z-v)^2}
\label{<Torb>}
\end{equation} 
where the conformal dimension of the twist $h$ is identified as: 
\begin{equation} 
h=\kappa/N=\frac{c}{24}\left(N-\frac{1}{N}\right).
\end{equation} \\

Note that we can now easily compute the simplest contribution $A^{\{0\}}=\int_{x=0}dz\left\langle \tilde{T}(z)\right\rangle _{\RN}$.  First one notices that in this expression,  $\tilde T$ can be replaced by $T$ since the term $\tilde T-T\propto \p Q_{\RN}$ is a total derivative and immediately
\begin{align}
A^{\{0\}} & =\int_{x=0}dz\frac{h}{N}\frac{(u-v)^{2}}{(z-u)^{2}(z-v)^{2}}=\frac{4i\pi h}{u-v}
\label{eq:valA0}
\end{align}
where one recalls that $u=-iL$ and $v=iL$, and where a factor of $N$ in the last expression comes from the $N$ disconnected part of the path of integration. An easy way to compute the integral in (\ref{eq:valA0}) consists in deforming the contour of integration in the half plane $\mbox{Im}(z)>0$ and pick up a residue at $z=v$.

\subsubsection{Higher order correlators}
\label{higherorders}

One can explicitly check the equivalence of the two routes when considering higher order correlators. Let us for example  compute the two-point function of $T_{\iorb}$  ; one gets
\begin{eqnarray}
\sum_{j_1,j_2}\vm{T_{j_1}(z_1)T_{j_2}(z_2)}_{\RN} &=& \sum_{j_1,j_2}\frac{(1-\xi_1)^4(1-\xi_2)^4}{(N(u-v))^4\;(\xi_1\xi_2)^2}\left[\kappa^2 + \frac{c}{2}\,\frac{(\omega^{(j_1)}_1\omega^{(j_2)}_2)^2}{(\omega^{(j_1)}_1-\omega^{(j_2)}_2)^4}\right]
\label{<TjTk>}\\
&=&\frac{1}{N^2}\frac{(1-\xi_1)^4(1-\xi_2)^4}{(\xi_1\xi_2)^2(u-v)^4}\left[
\kappa^2+\frac{c}{2}\;f_{2,2}(\omega^0_2/\omega^0_1)
\right]
\label{<TjTj>}
\end{eqnarray}
where one has written $\omega^{(j)}=e^{2i\pi j/N}\omega^0$, and one has introduced the function 

\begin{equation}
f_{p,q}(x)=\frac{x^q}{N^2}\sum_{j_1,j_2} \frac{e^{\frac{2i\pi}{N}(j_2-j_1)}}{(1-xe^{\frac{2i\pi}{N}(j_2-j_1)})^{p+q}}
=\delta(q)+ x^{-p}\frac{(-)^{p+q-1}}{(p\!+\!q\!-\!1)!}\;\p^{p+q-1}_{y}
\Big(\frac{y^{q-1}}{y^N\!-\!1}\Big)\Big|_{y=x^{-1}}.
\label{deffpq}
\end{equation}
The last equality shows that $f_{pq}$ is in fact an entire function of $x^N$ ; it can be derived by using an integral representation  $f_{pq}(x)=\frac{1}{2i\pi}\int_{\Cal C_{1^\pm}} \frac{dy}{y(y^N-1)}\frac{(xy)^q}{(1-xy)^{p+q}}$ where $\Cal C_{1^\pm}$ circles the unit circle, clockwise at radius $1-0^+$ and counter-clockwise at radius $1+0^+$. Moving the contour one gets Eq. (\ref{deffpq}),  so that $f_{2,2}(x)=N^3\left(\frac{x^N}{(1-x^N)^2}\right)^2+\frac{N(N^2-1)}{6}\;\frac{x^N}{(1-x^N)^2}$.
Note that, as it should, the final result (\ref{<TjTj>}) does not depend any longer on the sheet the points $z_1$ and $z_2$ are chosen in.
Putting everything together, one gets:
\begin{equation} 
\sum_{j_1,j_2}\vm{T_{j_1}(z_1)T_{j_2}(z_2)}_{\RN} = h^2 D^2+\frac{2hD}{(z_{12})^2}
+\frac{cN}{2(z_{12})^4}\mbox{\quad with } D=\textstyle \frac{(v-u)^2}{(z_1-u)(z_1-v)(z_2-u)(z_2-v)}
\label{<TT>}
\end{equation} 
This is again consistant with the Ward identity:
\begin{eqnarray}
\vm{T_{\iorb}(z_1)T_{\iorb}(z_2)\tw(u)\ttw(v)}&=&
\frac{\hat c}{2(z_{12})^4} \vm{\tw(u)\ttw(v)}\nonumber\\
&&\hspace{-3.5cm}+\left[ 
\textstyle
\frac{2}{(z_{12})^4}+\frac{\p_{z_1}}{z_{12}}+\frac{h}{(z-u)^2} + \frac{h}{(z-v)^2} +
\frac{\p_u}{z-u}+\frac{\p_v}{z-v}
\right]\;\vm{T_{\iorb}(z_2)\tw(u)\ttw(v)}
\end{eqnarray}

\subsection{Computations at finite size}
\label{twist@finite}

As we have seen in the section \ref{finitesize}, in order to compute the correlators on $\CN$, with the equivalence with the orbifold theory, one has to make another change of variables  from the plane to the cylinder 
$z=\frac{2L_{0}}{2\pi}\ln \omega$, that change the integration
contour and thus forbid to move it and wrap it around the twist fields, as we did in the infinite size case. For instance

\begin{align}
\sum_{i=1}^{N}\left\langle T(z^{(i)})\right\rangle _{\CN} & =\frac{\VMsurface[{\cal C}]{T_{\iorb}(z)\Phi(u)\tilde{\Phi}(v)}}{\VMsurface[{\cal C}]{\Phi(u)\tilde{\Phi}(v)}}\\
 & =\frac{\VMsurface[\mathbb{C}]{\left(\frac{dz}{d\omega}\right)^{-2}T_{\iorb}(\omega)\Phi(\tilde u)\tilde{\Phi}(\tilde v)}}{\VMsurface[\mathbb{C}]{\Phi(\tilde u)\tilde{\Phi}(\tilde v)}}\nonumber -\frac{\hat{c}}{12}\left(\frac{dz}{d\omega}\right)^{-2}\left\{ z;\omega\right\} 
\end{align}

And without too much pain, one can manage to compute the very first terms of the contributions
to the Renyi entropy, up to order $\mathcal{O}(\TB^{-2})$ (remember that $\partial\varphi^{(-k)}=\partial\varphi^{(k)*}$ and $\tilde v = e^{i \pi L/L_0} = \tilde u ^*$)

\begin{align}
\label{divergences}
A_{L_{0}}^{\{0\}} & =\frac{\pi }{L_{0}}\int_{0}^{\infty}d\omega
\left[ 
\frac{h\omega(\tilde u-\tilde v)^{2}}{\left|\omega-\tilde v\right|^{4}} -\frac{N}{24 \omega}\right]
=-\frac{2\pi h}{L}f_{1}(L/L_{0})- \underbrace{\int_0^\infty d\omega \frac{\pi N}{24 L_0 \omega}}_{div.}  
\end{align}

\begin{align}
A_{L_{0}}^{\left\{ 0;0\right\} } 
&= \left(A_{L_{0}}^{\{0\}}\right)^2 + \iint_0^\infty d\omega_1d\omega_2  \Big[
\underbrace{
\frac{h\pi^2\omega_1 \omega_2 }{L_0^2 (\omega_1-\omega_2)^2}
\frac{(\tilde u - \tilde v)^2 }{|\omega_1-\tilde v|^2 |\omega_2-\tilde v|^2 } }_{2h\pi^2 f_2(L/L_0)/L^2}
-\underbrace{\frac{N \pi ^2 \omega_1\omega_2}{4 L_0^2 (\omega_1-\omega_2)^4} }_{div.} \Big]
\nonumber\\
& =\frac{\pi^{2}}{L^{2}}2h\left\{ 2hf_{1}(L/L_{0})^{2}+f_{2}(L/L_{0})\right\} +div.
\label{eq:TNfinitesize}
\end{align}
where
\begin{align} 
f_{1}(a) &=a\left(1+\pi(1-a)\cot\pi a\right) \label{eq:f1}\\
f_{2}(a) & =f_1(a)^2+a^2(1-a)^2 \pi^2\label{eq:f2}
\end{align}

Note that these integrals are well defined, even in the presence of the $(\omega_i - \omega_j)^{-k}$ terms, because the integration points are "avoiding each other". This is a consequence of our point-splitting regularization scheme.

As discussed in the main text, the diverging terms in the integrals stemming from "intricated" integrands are proportional to $N$ only and are cancelled out by terms stemming from the denominator of Eq. (\ref{eq:renyientropy}).

\section{The composite field $\Cal O_4$}
\label{app-O4orb}

In this appendix we establish the operator product expansions of the first composite field $\Cal O_4^{\iorb}$ in the neutral sector of the orbifold theory with the twist operator $\tw$. The orbifold form of the perturbing operator is:
\begin{equation}
\Cal O_4^{\iorb} (z) \eveq \sum_{j=1}^N\big(\tilde T_j\tilde T_j\big)(z)\qquad;\quad \tilde T_j=T_j+\alpha \p Q_j
\label{defO4orb}
\end{equation}
where $z$ is a coordinate on the complex plane (punctured at points where the twist fields is inserted), and on the right hand side the index $j$ indicates the sheet of $\RN$. The parenthesis in the second line of (\ref{defO4orb}) indicates normal ordering $(A.B)(z)=\frac{1}{2i\pi}\oint_0 \frac{dx}{x} A(z+x) B(z)$ that is performed on $\RN$.

Introducing the Virasoro generators' action on a local field $O(z)$ via $L_{-k}(z)O(z) = \frac{1}{2i\pi}\oint_z dx\;(x-z)^{1-k}\;T_{\iorb}(x)O(z)$ \cite{belavinpolyakovzamolo84}, one parametrizes the OPE of $\Cal O_4^{\iorb}$ and $\tw$ as follows (only the descendents of the twist field can appear):
\begin{equation} 
\Cal O_4^{\iorb} (z)\cdot \tw(0)=\left[\textstyle \frac{\tilde a_4L_0}{z^4} + \frac{\tilde a_3L_{-1}}{z^3} +
\frac{\tilde a_2 L_{-1}^2+\tilde b_2L_{-2}}{z^2} + \frac{\tilde a_1L_{-1}^3+\tilde b_1L_{-1}L_{-2}+\tilde c_1 L_{-3}}{z}\right]\tw(0)
\label{OPEO4orb}
\end{equation} 

We will determine the coefficients $\tilde a_i,\tilde b_i,\tilde c_i$  (see Eqs.(\ref{coeffsO4orb}))  following the simplest  route: one starts by computing correlators 
$\Cal G_{X\Cal O}=\vm{X(z_i) \Cal O(z)}_{\RN}$ where $X(z_i)$ is an insertion of a collection of stress energy tensors $T_{\iorb}$ at points $z_i$, and one then expands this expression when $z\to u$. On the other hand,  one has $\Cal G_{X\Cal O}=\vm{\tw(u)\ttw(v)}^{-1}\vm{X(z_i)\Cal O(z)\tw(u)\ttw(v)}$: by identification this fixes constraints on the OPE $\Cal O(z)\cdot \tw(u)$ that are sufficient to determine the \eemph{a priori} unknown coefficients in Eq. (\ref{OPEO4orb}).

The operator (\ref{defO4orb}) can be split in 3 pieces, 
$$
\Cal O_4^{\iorb} \eveq (T^2)_{\iorb} + \alpha \left(\{\p Q,T\}\right)_{\iorb} + \alpha^2  \left(\p Q\p Q\right)_{\iorb}
$$
Let's start with $(T^2)_{\iorb}(z) \eveq \sum_{j}(T_jT_j)(z)= \oint_{z}\frac{dz_1}{z_1-z}\sum_{j}T_j(z_1)T_j(z)$. Note that the summation implies that $(T^2)_{\iorb}\neq (T_{\iorb})^2$.
Starting from the correlator (\ref{<TjTk>}), inserting $\delta(j_1-j_2)$ in the sum, and performing the normal order one finds:
\begin{equation} 
\sum_{j}\vm{(T_jT_j)(z)}_{\RN} = \frac{1}{N^3(u-v)^4}\frac{(1-\xi)^6}{\xi^4}\left[
\kappa^2(1-\xi)^2+\frac{\kappa}{60} P_1(\xi)
\right]
\label{<(TT)>}
\end{equation} 
with $P_1(X)=(1-X)^2(29N^2-11)+90N^2(1+X)^2$

Parametrizing the OPE of $(T^2)_{\iorb}$ with the twist field as:
\begin{equation} 
(T^2)_{\iorb} (z)\cdot \tw(0)=\left[\textstyle \frac{a_4L_0}{z^4} + \frac{a_3L_{-1}}{z^3} +
\frac{a_2 L_{-1}^2+b_2L_{-2}}{z^2} + \frac{a_1L_{-1}^3+b_1L_{-1}L_{-2}+c_1 L_{-3}}{z}\right]\tw(0),
\label{OPET2}
\end{equation} 
one can then expand the correlator  $\vm{(T^2)_{\iorb}(z)\tw(u)\ttw(v)}$ when $z$ approaches one of the twist fields, say $\ttw(v)$, and identify the two expansions obtained via the OPE (\ref{OPET2}) and via (\ref{<(TT)>}).
The first expansion involves correlators:
\begin{eqnarray}
\vm{\tw(u)\,L_{-n}\ttw(v)} &=& \frac{1}{2i\pi}\oint_v\frac{dz}{(z-v)^{n-1}} \vm{T_{\iorb}(z)\tw(u)\ttw(v)} = \frac{(n+1)h}{(u-v)^{n+2h}}\qquad (n\geq0)\nonumber\\
\vm{\tw(u)\, L_{-1}L_{-2} \ttw(v)} &=& \frac{6h(h+1)}{(u-v)^{3+2h}}
\end{eqnarray}
Now expanding (\ref{<(TT)>}) when $z\to u$, one gets the following identifications:
\begin{eqnarray}
a_4&=&\frac{1}{5cN}\left[h(22+5c)+9cN\right]\nonumber\\
a_3&=&\frac{1}{5cN}\left[2h(22+5c)+3cN\right]\nonumber\\
2(2h+1)a_2+3b_2&=&\frac{2h}{cN}(22+5c)\nonumber\\
(2h+1)(2h+2)a_1+(3h+1)b_1+2c_1&=&\frac{2h}{cN}(22+5c)
\end{eqnarray}
In order to remove the ambiguity in the determination of the unknown coefficients $a_i,b_i,c_i$, one needs to consider higher order correlator with more insertion of $T^{(0)}$:
\begin{eqnarray}
\sum_{j,k}\vm{T_j(z_1)(T_kT_k)(z_2)}_{\RN} &=& 
\frac{(1-\xi_1)^4(1-\xi_2)^8}{N^6(u-v)^6\;\xi_1^2\xi_2^4}
\sum_{j,k}\bigg[ 
\kappa^3 + 
\frac{\kappa c}{1440(1-\xi_2)^2}\,P_2(\xi_2) + \frac{\kappa c\;\omega_1^2\omega_2^2}{\omega_{12}^4}
\nonumber\\
&&+\;\;\frac{c}{12(1-\xi_2)^2}\sum_{p=0}^2\frac{\omega_1^2\omega_2^{4-p}}{\omega_{12}^{6-p}}Q_p(\xi_2)
\bigg]\nonumber\\
&=&\frac{(1-\xi_1)^4(1-\xi_2)^8}{N^4(u-v)^6\;\xi_1^2\xi_2^4}
\sum_{j,k}\bigg[ 
\kappa^3 + 
\frac{\kappa c\,P_2(\xi_2) }{1440(1-\xi_2)^2}+ \kappa c\;f_{2,2}\left(\frac{\omega_2^0}{\omega_1^0}\right)
\nonumber\\
&&+\;\;\frac{c}{12(1-\xi_2)^2}\sum_{p=0}^2 Q_p(\xi_2) f_{2,4-p}\left(\frac{\omega_2^0}{\omega_1^0}\right)  
\bigg]
\end{eqnarray}
with
\begin{eqnarray}
P_2(X)&=&
 (-22 - 100 N^2 + 122 N^4) X + (11 - 130 N^2 +        119 N^4) (1 + X^2)\nonumber\\
 Q_0(X)&=&36(1-X)^2\nonumber\\
  Q_1(X)&=&-36(1-X)[(N+1)X+N-1]\nonumber\\
   Q_2(X)&=& 5(1-X)^2 + 18N(X^2-1)+N^2(13X^2+10X+13)
\end{eqnarray} 
This fixes the coefficients $a_2,b_2$. A last constraint is needed: we compute the 3 points correlator $\sum_{j,k,l}\vm{T_j(z_1)T_k(z_2)(T_lT_l)(z_3)}_{\RN}$ (its expression is not particularly enlightening so we do not display it) and we take the limit $z_3\to v$. Finally  we find:
\begin{eqnarray}
a_1 &=& \frac{22+5c}{5 \hat{c}}\; \frac{-\hat{c}^2 + 23 \hat{c} h - 
 22 h^2}{(2 + \hat{c} - 7 h + \hat{c} h + 3 h^2) (\hat{c} - 10 h + 2 \hat{c} h + 
   16 h^2)} \nonumber\\
b_1 &=& \frac{22+5c}{5 \hat{c}}\; \frac{2 (\hat{c}^2 - 19 \hat{c} h + 3 \hat{c}^2 h + 54 h^2 - 21 \hat{c} h^2 + 2 \hat{c}^2 h^2 - 
   90 h^3 + 22 \hat{c} h^3 + 48 h^4)}{(2 + \hat{c} - 7 h + \hat{c} h + 
   3 h^2) (\hat{c} - 10 h + 2 \hat{c} h + 16 h^2)}\nonumber\\
c_1 &=& \frac{22+5c}{5 \hat{c}}\; \frac{2 (h-1) (\hat{c} - 12 h + \hat{c} h + 3 h^2)}{2 + \hat{c} - 7 h + \hat{c} h + 
   3 h^2}\nonumber\\
a_2 &=& \frac{22+5c}{5 \hat{c}}\; \frac{2 h (\hat{c} + 8 h)}{\hat{c} - 10 h + 2 \hat{c} h + 16 h^2} \nonumber\\
b_2 &=& \frac{22+5c}{5 \hat{c}}\; \frac{2 h (\hat{c} - 22 h + 2 \hat{c} h + 16 h^2)}{\hat{c} - 10 h + 2 \hat{c} h + 
   16 h^2}\nonumber\\
a_3 &=&\frac{3}{5}+ 2h\,\frac{22+5c}{5 \hat{c}}\nonumber\\
a_4 &=& \frac{9}{5}+ h\,\frac{22+5c}{5 \hat{c}}
\label{coeffsOPET2orb}
\end{eqnarray}
So far, the computation was totally general and applies to an arbitrary CFT put on a $\mathbb{Z}_N$. In our case, one deals with a free boson $Q\propto i\p\phi_+$, so that one should put $c=1$ and $\hat c=N$ in Eqs.(\ref{coeffsOPET2orb}).  

We are not done so far: there is a second piece to $\Cal O_4^{\iorb}$, namely terms that depend on the deformation parameter $\alpha$. On the plane, this deformation parameter simply amounts to a change of the central charge: $\tilde T$ is an energy-momentum tensor of a $c=1-6\alpha^2$ CFT. However, it would be \eemph{wrong} to identify $\Cal O_4^{\iorb}$ with $(\tilde T^2)_{\iorb}$, the orbifold version of the 
energy-momentum tensor of the deformed theory. Instead, what one should do is put the $c=1$ theory on the orbifold, and determine the orbifold version of the operators  $\alpha^2\sum_j \left(\p Q_j\p Q_j\right)$ and  $\alpha\sum_j  \left(\{\p Q_j,T_j\}\right)$. 

Consider  $\sum_j  \left(\p Q_j\p Q_j\right)$: the same game can be played with it, its OPE reading:
\begin{equation} 
 \left(\p Q\p Q\right)_{\iorb} (z)\cdot \tw(0)=\left[\textstyle \frac{\bar a_4L_0}{z^4} + \frac{\bar a_3L_{-1}}{z^3} +
\frac{\bar a_2 L_{-1}^2+\bar b_2L_{-2}}{z^2} + \frac{\bar a_1L_{-1}^3+\bar b_1L_{-1}L_{-2}+\bar c_1 L_{-3}}{z}\right]\tw(0)
\label{OPEdQdQ}
\end{equation} 
Proceeding in a similar way (we do not reproduce the steps, that are identical), we obtain the coefficients: $(\bar a_i,\bar b_i, \bar c_i) = -\frac{4}{9}(a_i,b_i,c_i)$, except for $\bar a_3=-\frac{4}{9} a_3+\frac{2}{3}$ and $\bar a_4=-\frac{4}{9} a_4+2$. Comparing this with the OPE:
\begin{equation} 
\p^2 T_{\iorb}(z) \cdot \tw(0) = \frac{6h\,\tw(0)}{z^4} + \frac{2L_{-1}\tw(0)}{z^3}, 
\end{equation} 
we find the remarquable relation:
\begin{equation} 
\sum_j\left[\left(\p Q_j\p Q_j\right) + \frac{4}{9}\left(T_j T_j\right)\right] = \frac{1}{3}\p^2 \sum_j T_j
\end{equation} 
so that keeping only terms that are even in $\alpha$ (terms linear in $\alpha$ are easily showed to be total derivatives, $(\{T,\p Q\})_{\RN} = \frac{2}{3}\p(TQ)_\RN$, and in general are odd under $Q\to -Q$, and  they don't contribute to  expectation values):
\begin{equation}
\Cal O_4^{\iorb}\Big|_{\alpha-\ind{even}} = \left(1-\frac{4\alpha^2}{9}\right)\;(T^2)_{\iorb} + \frac{\alpha^2}{3}\p^2T_{\iorb}
\label{dQdQTT}
\end{equation}
Restricting again to terms that are even in $\alpha$, one gets:
\begin{eqnarray}
\{\tilde  a_i,\tilde b_i,\tilde c_i\}_{i\leq 2} &=& \{\lambda a_i,\lambda b_i,\lambda c_i\} \nonumber\\
\tilde a_3 &=& \lambda\, a_3+2\alpha^2/3\nonumber\\
\tilde a_4 &=& \lambda\, a_4+2\alpha^2
\label{coeffsO4orb}
\end{eqnarray}
with $\lambda=1-4\alpha^2/9$, and $\{a_i,b_i,c_i\}$ are given by Eq.(\ref{coeffsOPET2orb}) with $c=1$.

\section{Computations with a direct quantization}
\label{directquantization}

We hereunder report some details useful in order to understand how to explicitly compute correlators with the mode decomposition.

\subsection{Energy density on the plane} 

With the mode decomposion it is easy to show that $\left\langle \varphi^{(k)}(\xi)\partial\varphi^{(-k)}(\xi^{\prime})\right\rangle =\sum_{m>0}\frac{1}{4\pi \xi^{\prime}}\left(\frac{\xi^{\prime}}{\xi}\right)^{m}$,
so that we find
\begin{align}
g^{(k)}(\xi,\xi^{\prime}) & =\left\langle \partial\varphi^{(k)}(\xi)\partial\varphi^{(-k)}(\xi^{\prime})\right\rangle \\
 & =-\frac{\left(\frac{\xi}{\xi^{\prime}}\right)^{\frac{k}{N}}}{4\pi(\xi-\xi^{\prime})^{2}}\left[(1-\frac{k}{N})+\frac{\xi^{\prime}}{\xi}\frac{k}{N}\right]\label{eq:correlator}
\end{align}
we then have the expression for the $k^{th}-$energy-momentum tensor
$\left\langle T^{(k)}(\xi)\right\rangle =-2\pi \left(g^{(k)}(\xi+\epsilon,\xi)+\frac{1}{\epsilon^{2}}\right)$
\begin{equation}
\left\langle T^{(k)}(\xi)\right\rangle =\frac{k(N-k)}{4N^{2}\xi^{2}}
\end{equation}
and summing up this expression over $k$ to get the total energy-momentum
tensor, we have
\begin{equation}
\left\langle T(\xi)\right\rangle =\frac{\hat{c}(N^{2}-1)}{24N^{2}}\frac{1}{\xi^{2}}
\end{equation}
with $\hat{c}=Nc$ $(c=1)$, since initially we have $N$ real bosonic
fields. This is the result that we would get doing the conformal transformation
$\omega=\xi^{1/N}$, this expression is the value of the schwartzian
derivative. The branch point in $\xi=0$ induces stress that leads to
a non vanishing expectation value for the energy-momentum tensor.

\subsection{Cut of finite size on the plane and on the cylinder}

If we are interested in a cut between two points $u,v$ on the $z-$plane,
all we have to do is make a conformal transformation that sends $v\rightarrow0$
and $u\rightarrow\infty.$ Namely
\begin{equation}
\xi=\frac{z-v}{z-u}\label{eq:transfoplanplan}
\end{equation}
since $\partial\varphi^{(k)}$ is a primary field, it is easy to compute
with the correlator (\ref{eq:correlator})
\begin{equation}
\left\langle T^{(k)}(z)\right\rangle ^{\{u,v\}}=\left(\frac{d\xi}{dz}\right)^{2}\left\langle T^{(k)}(\xi)\right\rangle^{ \{0,\infty\}} =\frac{k(1-k)(u-v)^{2}}{4N^{2}(z-u)^{2}(z-v)^{2}}
\end{equation}
and summing up over k, we find the expectation value of Eq. (\ref{eq:TNfinitesize})
\begin{equation}
\left\langle T(z)\right\rangle ^{\{u,v\}}=\frac{\hat{c}(N^{2}-1)}{24N^{2}}\frac{(u-v)^{2}}{(z-u)^{2}(z-v)^{2}}
\end{equation}
we see in the next section that we can find directly this result using
the energy momentum tensor's transformation identity for $T^{(k)}$
since the schwartzian of the transformation (\ref{eq:transfoplanplan})
is equal to zero.

Now we want to compute correlations on $\mathcal{C}_{N}$ : cylinders
of size $2L_{0}$ linked by a finite cut between $u=-iL$ and $v=iL$.
To come back to the usual situation where the fields acquire a phase
when turning around $z=0$, we have to make the following transformations
: first explode the cylinder on the plane, where $u=e^{-i\pi L/L_{0}},v=e^{i\pi L/L_{0}}$
and then transform $v\rightarrow0$ and $u\rightarrow\infty$, namely
it is given by
\begin{equation}
\xi=\frac{e^{\pi z/L_{0}}-e^{+i\pi L/L_{0}}}{e^{\pi z/L_{0}}-e^{-i\pi L/L_{o}}}\label{eq:transfoplancylindre}
\end{equation}
where $z\in\mathcal{C}.$

\subsection{Schwartzian}

It is interesting to check explicitly that even in this twisted geometry, at finite size,
and with this mode decomposition one finds that the schwartzian appears
when computing
\begin{align}
 & \left\langle T^{(k)}(z)\right\rangle _{cyl.}^{\{u,v\}}=\lim_{\epsilon\rightarrow0}-2\pi\left[\left\langle \partial\varphi^{(k)}(z+\epsilon)\partial\varphi^{(-k)}(z)\right\rangle _{cyl.}^{\{u,v\}}+\frac{1}{\epsilon^{2}}\right]\nonumber \\
 & =-2\pi\left[\frac{d\xi}{dz}(z)\frac{d\xi}{dz}(z+\epsilon)\left\langle \partial\varphi^{(k)}(\xi(z+\epsilon))\partial\varphi^{(-k)}(\xi(z))\right\rangle _{pl.}^{\{0,\infty\}}+\frac{1}{\epsilon^{2}}\right]\nonumber \\
 & =\frac{\pi^{2}k\left(1-k\right)e^{2\pi z/L_{0}}(u-v)^{2}}{4N^{2}L_{0}^{2}(e^{\pi z/L_{0}}-u)^{2}(e^{\pi z/L_{0}}-v)^{2}}-\frac{c\pi^{2}}{24L_{0}^{2}}
\end{align}
so that we will now directly use the well-known identity : $T'(z)=\left(\frac{d \xi}{dz}\right)^{2}T(\xi)+\frac{c}{12}\{\xi;z\}$,
even for the $T^{(k)}$ energy-momentum tensors.

\subsection{Example with $(\tilde{T}^{2})$}

Where the quantization scheme becomes really useful is when one is
interested in computing expectation values of operators that have
no simple expression in their orbifold version, for instance $(\tilde T^{2})$.

All we have to do is to use the mode decomposition \ref{modedecomposition} of the basic fields $\varphi_{(i)}$ and use the Wick theorem to compute explicitly 
\begin{align}
\int_{x=0}dz\left\langle (\tilde T^{2})(z)\right\rangle _{\mathcal{C}_{N}}=\sum_{i}\int_{x=0}dz\left\langle (T_{i}^{2})(z)\right\rangle _{cyl.}^{\{u,v\}} 
+\alpha^2
\underbrace{\sum_j \int dz\left\langle \left(\partial Q_j\partial Q_j\right)(z)\right\rangle _{\mathcal{C}_{N}}
}_{C}
\end{align}
where only the non vanishing expectation values have been written.
The integrand can be expressed with the fields $\partial\varphi^{(k)}$
as 
\begin{align} 
\sum_{j}\left\langle (T_{j}^{2})(z)\right\rangle _\CN 
& =\frac{1}{2i\pi N}\sum_{k,k^{\prime}}\oint_{\omega}\frac{d\epsilon}{\epsilon} 
 &&\Bigg[\underbrace{\left\langle T^{(k)}(z+\epsilon)\right\rangle \left\langle T^{(k^{\prime})}(z)\right\rangle }_{A} \nonumber\\
&&&
+\underbrace{2(-2\pi)^2\left\langle \partial\varphi^{(k)}(z+\epsilon)\partial\varphi^{(-k)}(z)\right\rangle \left\langle \partial\varphi^{(k')}(z)\partial\varphi^{(-k')}(z+\epsilon)\right\rangle }_{B}\Bigg]_\CN
\end{align}

\paragraph{Term A}
The term A expressed in the $\omega-$plane (change of variables \ref{CylToPlane}) is given by (we denote $a=L/L_{0}$) is
\begin{align}
A= &\int_0^\infty d\omega \Big[
\underbrace{
\frac{h^2\pi^3(\tilde u-\tilde v)^4\omega^3}
{L_0^3 N^3 |\omega-\tilde u|^8}}
_{}
-\underbrace{\frac{h\pi^3(\tilde v-\tilde u)^2 \omega}{12 L_0^3 N |\omega-\tilde u|^4 }}_{}
+\underbrace{\frac{N\pi^3}{576L_0^3\omega}}_{div.} \Big]\nonumber\\
= & \frac{h^2\pi^3 a^2}{3NL^3}\left(15\frac{f_1(a)}{\sin^2 a\pi}-6f_{1}(a)-5a\right) +\frac{h\pi^{3}}{6L^{3}}a^2f_1(a) + div.
\end{align}

\paragraph{Term B}  The term B gives 
\begin{align}
B=&\int_0^\infty d\omega \Big[
\frac{22 h^2\pi^3 (\tilde v-\tilde u)^4 \omega ^3}{ 5 N L_0^3  |\omega-\tilde u|^8}
+
 \frac{ h\pi^3 (\tilde u -\tilde v)^2\omega}{30 L_0^3 }  \left(\frac{25}{|\omega-\tilde u|^4} + \frac{90\omega(\tilde u+\tilde v)}{|\omega-\tilde u|^6} + \frac{54(\tilde u-\tilde v)^2\omega^2}{|\omega-\tilde u|^8} \right)
+\underbrace{\frac{11N\pi^3}{1440 L_0^3 \omega}}_{div.} \Big]
\nonumber\\
B=& \frac{22 h^2 \pi^3 a^2}{15NL^3} \left(15\frac{f_1(a)}{\sin^2 a\pi}-6f_{1}(a)-5a\right)
+\frac{11 h \pi^3}{15L^3}a^2 f_1(a) +div.
\end{align}

\paragraph{Term C}
And finally, the term C gives
\begin{align}
C= &  \int_0^\infty d\omega
\Big[
- \frac{12 h^2 \pi^3 (\tilde u -\tilde v)^4 \omega^3}{5 N L_0^3 |\omega-\tilde u|^8} + \frac{h\pi^3(\tilde u -\tilde v)^2 \omega}{5L_0 ^3} \left( \frac{5}{|\omega-\tilde u|^4} +\frac{10 \omega (\tilde u + \tilde v)}{|\omega-\tilde u|^6} +\frac{6(\tilde u -\tilde v)^2 \omega^2}{|\omega-\tilde u|^8} \right)
-\underbrace{\frac{N\pi^3}{240L_0^3\omega}}_{div.}
\Big] \nonumber\\
=& -\frac{4h^2\pi^3a^2}{5NL^3}\left(15\frac{f_1(a)}{\sin^2 a\pi}-6f_{1}(a)-5a\right)
 - \frac{ 2 h \pi^3}{5 L^3} a^2 f_1(a) +div.
\end{align}

The IR divergences  that come when integrating at large distance from the insertion of the twist operators will actually not depend on the twisted geometry. Hence these divergences will only get multiplied by a factor $N$ when one integrates on $\RN$, so that they will be compensated at every orders.  This is verified for the diverging terms stemming from A and B which are indeed linear in $N$, so that the expansion in power series of $\mathcal{O}(1/L\TB)$ of the exponential $\exp(-\int_\RN dz A_{div}^{\{1\}})$ will be exactly compensated from terms of the denominator $\exp(-N\int_{\mathcal{R}_1} dz A_{div}^{\{1\}})$ in the expression of $R_{N}$.

Combining the results one finally gets ($a=L_0/L$)
\begin{align}
A_{L_0}^{\{1\}} =\frac{(9-4\alpha^2)}{5}\left[\frac{3h^2\pi}{NL^3} f_4(a) +\frac{h\pi^3a^2f_1(a)}{2L^3}\right]
\end{align}
where
\begin{align}
f_4(a)&=\frac{\pi^2a^2}{15} \left(15f_1(a)/\sin^2(a\pi) - 6f_1(a) -5a\right) \\
&=1-2\pi^2a^2/5+\pi^4a^4/15+\mathcal{O}(a^6)
\end{align}

It is easy to check the result known at infinite size for $A^{\{1\}}$ (from Eq. (\ref{resultsAi})): 
\begin{align}
A^{\{1\}}=\frac{3(9-4\alpha^2)h^2\pi}{5NL^{3}}=-\frac{3\pi h^2 (4-D)(1-4D)}{DNL^3}
\end{align}

Note that now, $A_{L_0}^{\{1\}}$ gives non-vanishing corrections to
the entropy (terms proportional to $h$). These corrections are due to finite size effects, they are proportional to $a^2$ and depend on the interactions ($\alpha^2$ term).



\end{document}